\documentclass[apjl]{emulateapj}

\usepackage{graphicx,natbib}
\usepackage{amsmath}
\usepackage{apjfonts}

\newcommand{\bootes}{Bo\"otes}
\newcommand{\bootess}{Bo\"otes }

\newcommand{\s}{$\sim$}
\newcommand{\muu}{$\mu$}

\begin{document}

\author{
Richard Beare\altaffilmark{1, 5},
Michael J. I. Brown\altaffilmark{1},
Kevin Pimbblet\altaffilmark{2, 1},
Fuyan Bian\altaffilmark{3},
Yen-Ting Lin\altaffilmark{4},
}

\altaffiltext{1}{Monash Centre for Astrophysics, School of Physics and Astronomy, Monash University, Clayton, Victoria 3800, Australia. }
\altaffiltext{2}{Department of Physics and Mathematics, University of Hull, Cottingham Road, Kingston-upon-Hull, HU6 7RX, UK.}
\altaffiltext{3}{Stromlo Fellow, Mount Stromlo Observatory, Research School of Astronomy and Astrophysics, The Australian National University, Cotter Road, Weston, ACT 2611, Australia.}
\altaffiltext{4}{Institute of Astronomy and Astrophysics, Academia Sinica, Taipei 10617, Taiwan}
\altaffiltext{5}{Email: richard@beares.net}

\slugcomment{Accepted for publication in the Astrophysical Journal}

\date{\today}

\title{THE $z < 1.2$ OPTICAL LUMINOSITY FUNCTION FROM A SAMPLE OF $\sim410 \, 000$  GALAXIES IN BOOTES}

\shorttitle{OPTICAL LUMINOSITY FUNCTION EVOLUTION}

\shortauthors{Beare \it {et al.}}

\begin{abstract}

Using a sample of $\sim410 \, 000$  galaxies to depth $I_{\textrm{AB}} = 24$ over 8.26 deg$^2$ in the \bootess field ($\sim 10$ times larger than $z \sim 1$ luminosity function studies in the prior literature), we have accurately measured the evolving $B$-band luminosity function of red galaxies at $z<1.2$ and blue galaxies at $z<1.0$. In addition to the large sample size, we utilise photometry that accounts for the varying angular sizes of galaxies, photometric redshifts verified with spectroscopy, and absolute magnitudes that should have very small random and systematic errors. Our results are consistent with the migration of galaxies from the blue cloud to the red sequence as they cease to form stars, and with downsizing in which more massive and luminous blue galaxies cease star formation earlier than fainter less massive ones. Comparing the observed fading of red galaxies with that to be expected from passive evolution alone, we find that the stellar mass contained within the red galaxy population has increased by a factor of \s3.6 from $z \sim 1.1$ to $z \sim 0.1$. The bright end of the red galaxy luminosity function fades with decreasing redshift, the rate of fading increasing from \s0.2 mag per unit redshift at $z = 1.0$ to \s0.8 at $z = 0.2$. The overall decrease in luminosity implies that the stellar mass in individual  highly luminous red galaxies increased by a factor of \s2.2 from $z = 1.1$ to $z = 0.1$. 

\end{abstract}

\keywords{galaxies: abundances - galaxies: evolution -  galaxies: statistics.}

\section{Introduction}
\label{sec:LFintro}

Galaxy luminosity is a function of stellar mass and star formation history and thus provides an indirect measure of the build up of stellar mass within galaxies. Although the galaxy stellar mass function (SMF) is more fundamental than the luminosity function (LF), it cannot be measured directly as the LF can. Galaxy stellar masses can be deduced from galaxy luminosities if stellar mass to light ($M/L$) ratios are known, or they can be measured by fitting stellar population synthesis (SPS) models directly to observed photometry, but either way considerable uncertainties are present in regard to the models used. LFs on the other hand can be measured directly from from observed photometry with smaller uncertainties. There is therefore an important role for accurate measurement of the evolution of LFs, which can be compared directly with theoretical models of galaxy growth and evolution, placing key constraints on their assumptions and parameters. For passive galaxies, stellar $M/L$ ratios are reasonably well defined and reasonably accurate measurements of red galaxy stellar mass growth can be inferred directly from evolution of the LF.

Studies of LF evolution have used both spectroscopic and photometric redshifts. Spectroscopic surveys provide precision galaxy redshifts, but have smaller sample sizes and smaller volumes, and require larger telescopes than comparably deep photometric redshift surveys. However, photometric redshifts can suffer from catastrophic errors, so that purely imaging surveys can suffer from large systematic errors.  Key aspects of several recent spectroscopic and photometric studies of LF evolution and LFs in the low redshift ($z < 0.2$) Universe are summarised in Table \ref{tab:LF_literature}. A useful review of past measurements of the optical LF is given by \citet{johns11}.

\begin{deluxetable*}{ccccccc}						
\tablewidth{0pt}						
\tablecolumns{7}						
\tabletypesize{\scriptsize}						
\tablecaption {Recent key measurements of optical luminosity function and its evolution}						
\tablehead{						
\colhead{reference} & 	\colhead{surveys} & 	\colhead{redshift range} & 	\colhead{redshift} & 	\colhead{faint} & 	\colhead{sample} & 	\colhead{approx.} \tabularnewline 
\colhead{} & 	\colhead{used} & 	\colhead{in survey} & 	\colhead{type} & 	\colhead{limit} & 	\colhead{size} & 	\colhead{sample area} \tabularnewline 
\colhead{} & 	\colhead{} & 	\colhead{} & 	\colhead{(s or p)} & 	\colhead{(AB)} & 	\colhead{} & 	\colhead{(deg$^2$)} \tabularnewline 
}						
 \startdata						
\\						
\multicolumn{7}{l}{LOW REDSHIFT STUDIES}\\						
\\						
\citet{blant01} & 	SDSS (commissioning) & 	$<0.2$ & 	s &	$r^*=17.6$ &	$11\,273$ & 	$140$ \\[4pt]
\citet{norbe02} & 	2dFGRS & 	$z<0.25$ & 	s &	$b_J=19.45$ &	$115\,986$ & 	$\sim2\,000$ \\[4pt]
\citet{madgw02} & 	2dFGRS & 	$0.01<z<<0.15$ & 	s &	$b_J=19.45$ &	$75\,589$ & 	$\sim2\,000$ \\[4pt]
\citet{blant03} & 	SDSS & 	$0.02 < z < 0.22$ & 	s &	$^{0.1}r=17.79$ &	$147\,986$ & 	$1\,844$ \\[4pt]
\citet{blant05} & 	SDSS DR2 (VAGC) & 	$z < 0.05$ & 	s &	$^{0.1}r=17.77$ &	$28\,089$ & 	$2\,221$ \\[4pt]
\citet{blant06} & 	SDSS DR4 (VAGC) & 	$0.05 < z < 0.15$ & 	s &	$^{0.1}r=17.6$ &	$\sim430\,000$ & 	$2\,627$ \\
- & 	 + DEEP2 & 	$0.8 < z < 1.2$ & 	s &	$R=24.31$ &	$2\,976$ & 	$0.63$ \\[4pt]
\cite{monte09} & 	SDSS DR6 (VAGC) & 	$0.02 < z < 0.22$ & 	s &	$r=17.6$ &	$437\,565$ & 	$7\,280$ \\[4pt]
\citet{hill10} & 	SDSS + MGC-Bright + UKIDSS & 	$0.0033 < z < 0.1$ & 	s &	$r = 17.9$ &	$2\,781$ & 	$30.88$ \\[4pt]
\citet{loved12} & 	GAMA DR1 & 	$0.02 < z < 0.5$ & 	s &	$r=19.45$ &	$12\,789$ & 	$144$ \\
\\						
\multicolumn{7}{l}{STUDIES OF LF EVOLUTION}\\						
\\						
\citet{wolf03} & 	COMBO17 & 	$0.2 < z < 1.2$ & 	p &	$R=24.31$ &	$25\,000$ & 	$0.78$ \\[4pt]
\citet{bell04} & 	COMBO17 & 	$0.2 < z < 1.2$ & 	p &	$R=24.31$ &	$25\,000$ & 	$0.78$ \\[4pt]
\citet{loved04} & 	SDSS DR1 & 	$0.001 < z < 0.3$ & 	s &	$r=17.6$ &	$90\,000$ & 	$2099$ \\[4pt]
\citet{ilber05} & 	VVDS & 	$0.05 < z < 2.0$ & 	s &	$I=24$ &	$7\,840$ & 	$0.61$ \\[4pt]
\citet{ilber06b} & 	VVDS + COMBO17 + HST/ACS & 	$0.05 < z < 1.2$ & 	s, p &	$I=24$ &	$605, 3555$ & 	$0.044$ \\[4pt]
\citet{zucca06} & 	VVDS & 	$0.05 < z < 1.5$ & 	s &	$I=24$ &	$7\,713$ & 	$0.61$ \\[4pt]
\citet{wake06} & 	SDSS + & 	$0.17 < z < 0.24$ & 	s &	$r=17.6$ &	$6326$ (LRGs) & 	$180$ \\
- & 	 + 2SLAQ & 	$0.5 < z < 0.6$ & 	s &	$i < 19.8$ &	$1725$ (LRGs) & 	$180$ \\[4pt]
\citet{willm06} & 	DEEP2 & 	$0.2 < z < 1.4$ & 	s &	$R=24.31$ &	$11\,000$ & 	$1.13$ \\[4pt]
\citet{faber07} & 	DEEP2 & 	$0.2 < z < 1.2$ & 	s &	$R=24.31$ &	$11\,000$ & 	$1.13$ \\
- & 	 + COMBO17 & 	$0.2 < z < 1.2$ & 	p &	$R=24.21$ &	$39\,000$ & 	$1.9$ \\[4pt]
\citet{brown07} & 	NDWFS + SDWFS & 	$0.2 < z < 1.0$ & 	p &	 $I=23.95$ &	$39\,599$ (red) & 	$7.0$ \\[4pt]
\citet{zucca09} & 	zCOSMOS-bright + HST ACS & 	$0.1 < z < 1.0$ & 	s &	$I=22.5$ &	$10\,644$ & 	$1.4$ \\[4pt]
\citet{cool12} & 	AGES & 	$0.05 < z < 0.75$ & 	s &	$I=20.4$ &	$12\,500$ & 	$7.6$ \\[4pt]
\citet{loved12} & 	GAMA DR1 & 	$0.002 < z < 0.5$ & 	s &	 $r=19.4$ &	$90\,000$ & 	$144$ \\[4pt]
\citet{fritz14} & 	VIPERS & 	$0.4 < z < 1.3$ & 	s &	$i=22.5$ &	$45\,000$ & 	$10.32$ \\[4pt]
this work & 	NDWFS + NEWFIRM + SDWFS & 	$0.2 < z < 1.2$ & 	p &	 $I=24.0$ &	$408 \, 495$ & 	$8.26$ \\
\enddata						
\label{tab:LF_literature}						
\end{deluxetable*}

Passive and star-forming galaxies have most commonly been distinguished on the basis of a red/blue cut in restframe color-magnitude space, and most $z> 0.2$ studies \citep[e.g.][]{bell04, willm06, faber07, brown07} have used an evolving cut to model the way that both the red sequence and the blue cloud have become redder with time. A number of other studies have used morphological indices to differentiate different galaxy types, \citep[e.g. bulge-dominated/disk-dominated/irregular,][]{ilber06, zucca09}. Others have used broadband spectrophotometric indices, \citep[e.g.][]{madgw02, zucca09}. Restframe color, morphology and spectral type are all   intended as proxies for distinguishing quiescent from star-forming galaxies, but several authors point out that many low luminosity red galaxies are in fact dusty edge-on spirals \citep[e.g.][]{weine05, willi09, bell12, dolle14},  while recent work has also shown that many spiral galaxies are in fact red in color \citep[e.g.][]{bonne15}. Interpretations of LF evolution must take this uncertainty into consideration.

Luminosity functions are generally parameterised using a Schechter function:

\begin{multline} \phi_M \left( M \right) dM \label{eq:schechter_M}
	\\
	= 0.4 \ln{10} \phi^* 10^{-0.4 (\alpha + 1) (M - M^*) } \exp ({ -10^{-0.4(M - M^*) } } )dM
\end{multline}\\
where $\phi^*$ is a normalising factor, $M$ is the absolute magnitude in a given waveband, $M^*$ corresponds roughly to the transition from a power law luminosity function to an exponential one, and $\alpha$ determines the slope of the power law variation at the faint end.  Because $\alpha$ becomes harder and harder to determine as redshift increases, most high redshift studies have used fixed values of $\alpha$ equal to those measured in the lowest redshift bins. 

Several low redshift i.e. ($z \lesssim 0.2$) studies have found that there is an excess of very faint red galaxies above the number that can be modelled using a simple Schechter function, and they have accordingly added additional terms to the Schechter function to model this \citep[e.g.][]{madgw02, blant06, loved12}. In common with other measurements of LF evolution we do not reach sufficiently faint restframe magnitudes for these modifications to be significant for our work.

Most studies agree that for blue galaxies the space density parameter $\phi^*$ has remained roughly constant since $z = 1$ \citep[e.g.][]{wolf03, willm06, faber07}. For quiescent/red galaxies, the majority of studies agree that $\phi^*$ has increased with decreasing redshift, but they give widely differing estimates of the factor by which it has increased.  Similarly, all authors agree that the characteristic magnitudes $M^*$ of galaxies have become fainter with time, but there are considerable differences for the same waveband in the estimates as to how much fainter, and differences as to whether $M^*$ for red or $M^*$ for blue galaxies fades faster. Much of this variation in measured $\phi^*$ and $M^*$ values can be attributed to the highly degenerate nature of the three Schechter parameters, with the adopted value of $\alpha$ making a significant difference to the other two parameters (as we discuss later in \S\ref{sec:literature} in regard to our own results).  

The luminosity density measures the amount of light produced by a galaxy population and is thus an indicator of the stellar mass within that galaxy population. Its evolution has been used to infer the growth of stellar mass within the red galaxy population. For an LF given by a Schechter function, the luminosity density is:

\begin{equation} \label{eq:lumdens}
	j = \phi^* L^* \Gamma(\alpha + 2).
\end{equation}

where $L^*$ is the luminosity corresponding to the characteristic magnitude $M^*$.

Fortunately measurements of luminosity density  $j$ vary much less than those of the Schechter parameters $\phi^*$ and $M^*$ because decreased $M^*$ estimates (brighter luminosities) correlate with smaller $\phi^*$ estimates. Furthermore, for red galaxies (for which $\alpha \sim -0.5$), $j$ varies little with $\alpha$ because $\Gamma(\alpha + 2)$ has a local minimum at $\alpha=-0.5$.  There is overall agreement in the literature that the luminosity density of blue galaxies has decreased since $z=1$ while that of red galaxies has changed little. 

Because the passive fading of quiescent galaxies can be modelled using stellar population synthesis (SPS) models \citep[e.g.][]{bruzu03}, a number of authors have been able to draw conclusions regarding the build up of stellar mass within the red galaxy population. For example, \citet{bell04} and \citet{brown07} both estimated that the stellar mass within red galaxies has doubled since $z=1$. A number of authors have inferred stellar mass function evolution from optical LF evolution  \citep[e.g.][]{bell03, taylo09} and from near infrared LF evolution  \citep[e.g.][]{bundy06, borch06, ilber10} by using stellar mass to light (M/L) ratios derived from theoretical models.

An additional measurement derived from LF evolution results by some authors \citep[e.g.][]{bell04, brown07} is how the most luminous red galaxies (LRGs) have changed in luminosity over time. \citet{bell04} used an argument based on SPS models to demonstrate that there were insufficient massive blue galaxies at $z \sim 1$ to produce today's luminous red galaxies when they ceased to form stars. These LRGs must therefore have grown in stellar mass and luminosity by mergers with smaller ellipticals or by dusty mergers in which any bursts of star formation are obscured by dust. They estimated that the stellar mass in individual LRGs has doubled since $z=1$, while \citet{brown07} concluded that 80$\%$ of it was already in place at $z=0.7$. These authors measured luminosity function evolution for highly luminous galaxies by determining how the absolute magnitude at constant space density has evolved. This can be done for any galaxy sample for which the Schechter function parameters have been determined and we make this additional calculation later for a number of studies (\S\ref{sec:results_luminous}).

\citet{zucca09} investigated the role of environment on the evolution of different types of galaxy. They divided their sample by both morphology (E + S0, spiral, irr) and by spectrophotometric type, and concluded that the bulk of the transformation from blue galaxies to red probably happened before $z \sim 1$ in overdense regions, but was still ongoing at lower redshifts in underdense environments. Galaxies in ``overdense'' and ``underdense'' regions were defined to be those in the upper and lower quartiles of the overdensity distribution when overdensity was computed using a 5th nearest neighbour estimator.

Given the considerable variation in conclusions drawn from the various studies summarised in Table \ref{tab:LF_literature}, there is clearly a need for additional more accurate measurements of optical LF evolution, particularly at $z\ga0.5$. Published studies of the luminosity function using large samples to $z\sim1$ are those based on COMBO17 \citep{wolf03, bell04}, DEEP2 \citep{willm06}, zCOSMOS \citep{zucca09} and VIPERS \citep{fritz14}, as well as that of \citet{brown07} for red galaxies using NDWFS and SDWFS data. COMBO17 and DEEP2 are compared in \citet{faber07}. Although the combined sample of \citet{faber07} numbers $39\,000$ galaxies covering an area of nearly 2 deg$^2$, the combination of cosmic variance and Poisson statistics still produces an uncertainty in $\phi^*$ of \s14$\%$.

In this paper we take advantage of the very large sample size available in \bootess (an order of magnitude greater than any previous survey): $408\,495$ galaxies over 8.26 deg$^2$ measured to a depth of $I = 24.0$ (AB) in several optical and near infrared wavebands and use this to measure evolution of the $B$-band optical luminosity function over the range $0.2 < z < 1.2$. Our work is an extension of that by \citet{brown07}, using improved photometry and including blue galaxies as well as red and an extra redshift bin ($1.0 \leq z < 1.2$). We also make use of the newly available atlas of 129 accurate empirical galaxy SEDs from \citet{brown14} to determine accurate photometric redshifts and accurate absolute magnitudes using the method of \citet{beare14}. This paper is the first of two based on the \bootess data. Paper II measures evolution of the $K$-band LF and then uses both optical and infrared stellar mass to light ratios to measure evolution of the galaxy stellar mass function.

The structure of this paper is as follows: Section \ref{sec:surveys} describes the surveys that we have used,  Section \ref{sec:apertures} describes object detection and photometry, Section \ref{sec:photoz} describes measurement of photometric redshifts, Section \ref{sec:selection} describes sample selection, Section \ref{sec:absmags} explains how we calculated absolute magnitudes from our photometry and Section \ref{sec:lumfn} describes determination of LFs. We present our results and discuss them in Section \ref{sec:results}. Finally we summarise our work and conclusions in Section \ref{sec:summary}.

Our results are determined assuming a cosmology with $\Omega_0 = 0.3$, $\Omega_{\textrm{k}}=0$, $H_0=70 \, \textrm{km s}^{-1} \,  \textrm{Mpc}^{-1}$ which is similar to that implied by WMAP measurements \citep{benne13}, and presented using AB-based magnitudes and units in which $h_{70} = H_0 /70$. Conversions to other cosmologies can be made as described in \citet{croto13}.

\section{The surveys}
\label{sec:surveys}

We used data from several legacy surveys covering 8.26 square degrees in \bootess to determine photometric redshifts, to calculate absolute (restframe) magnitudes, to apply various color cuts, and to separate red and blue galaxies on the basis of restframe color bimodality.  Our photometry is based on $B_{\rm{W}}$, $R$ and $I$-band images from the third data release of the NOAO Deep Wide Field Survey \citep[NDWFS,][]{jannu99}, $J$, $H$ and $K_{\rm{S}}$-band images from the NEWFIRM  \bootess Imaging Survey \citep{gonzainprep}, $u$ and $y$-band images from the $2 \times 8.4$ m Large Binocular Telescope \citep[LBT;][]{bian13}, $z$-band data from the 8.2 m Subaru Telescope \citep{miyaz12}, and 3.6, 4.5, 5.8 and 8.0 $\mu$m near infrared images from the Spitzer Deep Wide Field Survey \citep[SDWFS;][]{ashby09, eisen08}. $11\,087$ spectroscopic redshifts from several sources (Section \ref{sec:specz}) were used in preference to photometric redshifts when available ($3.0 \%$ of the total redshifts). They were also used to verify our photometric redshifts (Section \ref{sec:photoz}).

\subsection{NOAO Deep Wide Field Survey (NDWFS)}
\label{sec:NDWFS}

The NOAO Deep Wide Field Survey imaged two fields of approximately 9.3 square degrees each, one in \bootess using the MOSAIC-I camera on the KPNO 4 metre telescope, and one in Cetus using multiple instruments and telescopes.    The NDWFS $5 \sigma$ (AB) magnitude detection limits are $B_{\rm{W}} = 26.6$, $R = 26.0$ and $I = 26.0$. 

\subsection{Spitzer Deep Wide Field Survey}
\label{sec:SDWFS}

We used 3.6, 4.5, 5.8 and 8.0 micron  infrared photometry from the Spitzer Deep Wide Field Survey \citep[SDWFS;][]{ashby09, eisen08} Infrared Array Camera \citep[IRAC;][]{fazio04} in determining photometric redshifts and for color cuts to exclude stars and AGN. Average $5 \sigma$ (AB) depths in these wavebands were 22.6, 22.1, 20.3 and 20.2 respectively.

\subsection{NEWFIRM \bootess Imaging Survey}
\label{sec:NEWFIRM}

$J$,  $H$ and $K_{\rm{S}}$-band data from Data Release 2 of the NEWFIRM \bootess Imaging Survey were used for the photometric redshifts and the $J$-band data for photometry.  

The NEWFIRM survey \citep{autry03} covered the whole of the \bootess region covered by the NDWFS and SDWFS surveys and made use of the NOAO Extremely Wide-Field Infrared Imager (NEWFIRM camera) on the Mayall 4 metre telescope on Kitt Peak.  The survey reached $5 \sigma$ (AB) depths of at least $J = 22.9$, $H = 22.1$ and $K_{\rm{S}} = 21.3$ within a 3 arcsecond diameter aperture.

\subsection{The LBT \bootess Field Survey}
\label{sec:LBT}

Imaging from the LBT Large Binocular Cameras \citep[LBCs;][]{giall08} in the $u$ and $y$ bands was used for the photometric redshifts. The $2 \times 8.4$ m LBT was used in binocular mode, with the two LBCs imaging the same region of \bootess in $u$ and $y$ simultaneously. Each portion of the \bootess field was observed for approximately 1200 seconds, with 240 second individual exposures and a $30^{\prime\prime}$ dithering pattern being used to fill gaps between the LBC CCDs. The LBT survey 5$\sigma$ (AB) magnitude detection limits were $u=25.2$ and $y=24.4$.  We refer the reader to  \citet[][]{bian13} for a more thorough description of the survey.

\subsection{Subaru $z$-band imaging}
\label{sec:Subaru}

$z$-band imaging from the SuprimeCam camera on the 8.2 m Subaru telescope was used for the photometric redshifts \citep[][]{miyaz12}. Almost the whole \bootess field was imaged using exposure times of either 12 or 24 minutes, and the majority of the 39 pointings resulting in images with seeing of 0.7 arcsec or better.  Data reduction was carried out with the prototype pipeline for the Hyper SuprimeCam (HSC) and the photometry was calibrated to the Sloan Digital Sky Survey \citep{york00}. The 5$\sigma$ AB magnitude detection limit (detected with a 3 arcsec diameter aperture) was $z\sim24.1$.

\subsection{Spectroscopic redshifts}
\label{sec:specz}

The vast majority of spectroscopic redshifts we were able to use in the \bootess field came from the AGN and Galaxy Evolution Survey \citep[AGES,][]{kocha11}, which obtained spectra of $18\,163$ galaxies with $I$-band magnitudes brighter than 20.5 out to $z = 1$, (and quasars with $I< 22.0$ out to redshift 6.5).  AGES used the Hectospec Multiobject Optical Spectrograph on the 6.5 metre MMT telescope at Mount Hopkins. Several hundred additional redshifts were obtained from SDSS and from a variety of programs with the Gemini, Keck and Kitt Peak National Observatory telescopes. 

\section{Object detection and photometry}
\label{sec:apertures}

Copies of the $B_wRIyHK_s$ images and the four IRAC band images were smoothed to a common Moffat point spread function with a full width at half-maximum (FWHM) of $1.35''$. $u$, $z$ and $J$ images were smoothed to give FWHM values of $1.60''$, $0.68''$ and $1.60''$ respectively. These FWHM values were chosen to correspond to the image with the worst seeing.  This ensured that the fraction of the light captured by small apertures did not vary from filter to filter and from subfield to subfield across the \bootess field.

We used a similar galaxy catalog to \citet{brown08} and sources were detected using SExtractor 2.3.2 \citep{berti96} run on unsmoothed $I$-band images from the NWDFS third data release.  Duplicate object detections were removed from the small regions of overlap between subfields.    To minimize contamination of the catalogue, regions surrounding very extended galaxies and saturated stars were flagged and excluded from the analysis. Visual inspection confirmed that the majority of these regions did in fact surround saturated stars or bright galaxies.  Fifty-five bright galaxies with SDSS spectroscopic redshifts lay within the excluded regions and we added these back in so that the bright end of the luminosity function was not biased.  Of these, 36 lay in the first redshift bin $0.2 \leq z < 0.4$.  The final sample covered an area of 8.262 deg\({}^2\) over a \( 2.9^\circ \times 3.6^\circ \) field of view.

\citet{brown08} used their own code to measure the apparent magnitude of each source in each waveband using apertures with diameters ranging from 1 to 20 arcsecond.  SExtractor segmentation maps were used to exclude flux associated with neighbouring objects.  Corrections were also made for missing pixels (e.g. bad pixels) using the mean flux per pixel measured in a series of annuli surrounding each object.   Random uncertainties were estimated by measuring the flux at $\simeq 100$ or so positions near to each detected object. As described in \citet{brown07}, they verified the uncertainty estimates  using artificial galaxies with \citet{dev48} profiles that were added to copies of the data and measured with the photometry code.

\begin{figure}
 	\centering
		\includegraphics[width=0.45\textwidth]{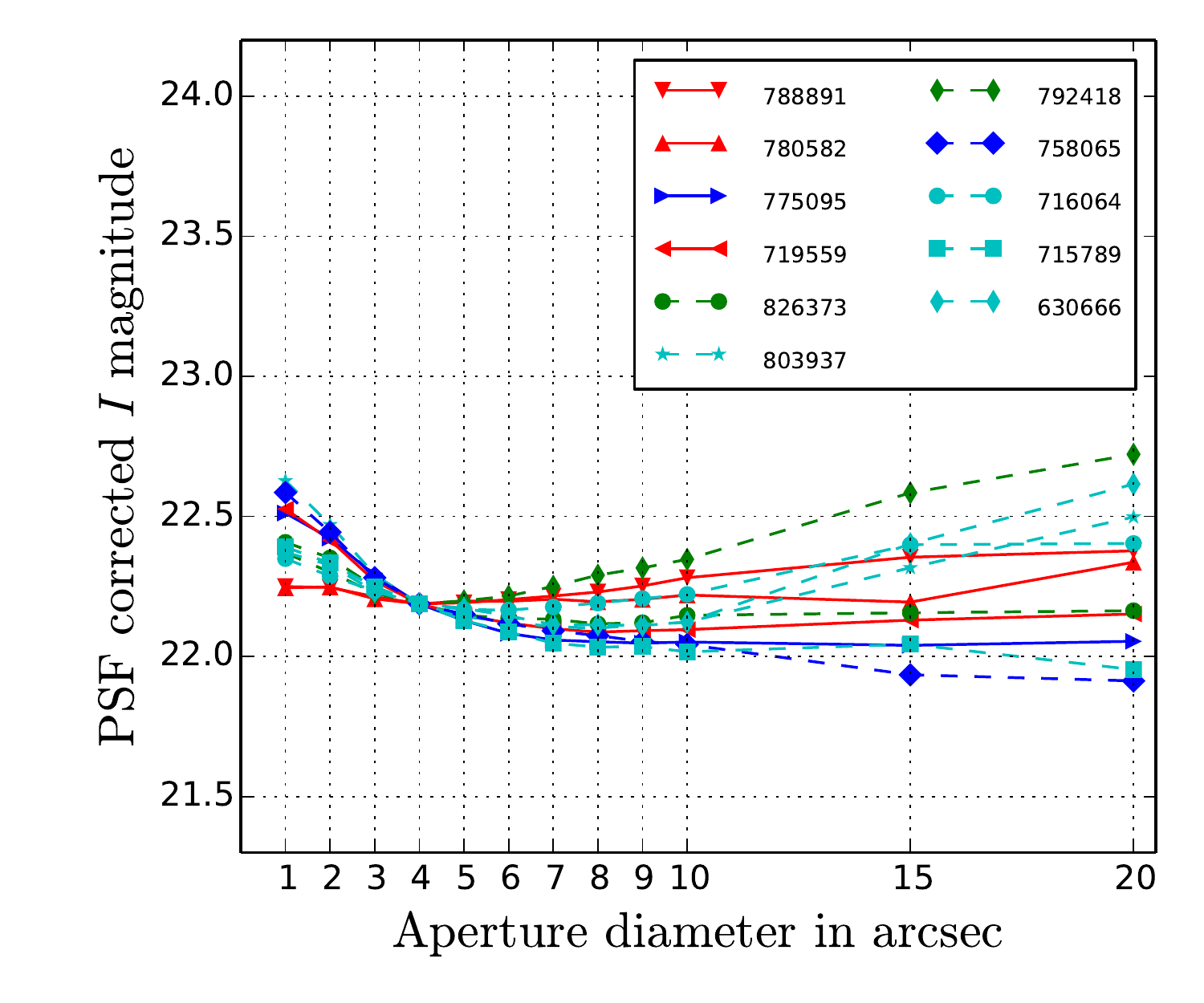}		
		\caption{Example PSF-corrected growth curves showing how the optimum photometric aperture diameter for $22.0 \leq I < 23.0$ was determined. The curves are based on uncorrected 4 arcsec magnitudes and have been normalised to a common 4 arcsecond magnitude. Solid lines are for galaxies with no significant near neighbours and dashed lines for galaxies with only faint or marginal contamination. Red, green, cyan and blue correspond to the four redshift bins from $0.2\leq z<0.4$ to $0.8\leq z<1.0$.}
		\label{fig:growth_curves}
\end{figure}

We employed a variable aperture diameter between 3 and 15 arcsecond, dependent on the $I$-band magnitude measured using a 4 arcsecond diameter aperture.  We identified galaxies in different apparent magnitude ranges which appeared from visual inspection to have no near neighbours. This was done by overlaying concentric circles of differing diameters around individual galaxy images displayed using the image visualisation tool ds9.  Using just these isolated galaxy images, we then plotted Moffat point spread function (PSF) corrected magnitudes as a function of aperture diameter  as shown in Figure \ref{fig:growth_curves}, and selected an aperture where the magnitude as a function of aperture diameter changed (on average) by less than 0.03  $\rm{\, mag \, arcsecond}^{-1}$. In the case of galaxies with $I>21.0$, we used an area approximately 50\% smaller than that obtained by the preceding procedure, so that we avoided including any small amounts of extraneous light which would be proportionately more significant for these fainter objects.  We then normalised the growth curves to the chosen aperture diameter and calculated the total correction as the sum of a Moffat PSF correction and the mean offset at larger apertures for the normalised growth curves.  Table \ref{tab:apertures} lists the apertures we used for all wavebands except $J$, for which slightly larger corrections were used due to the broader PSF in this waveband.

To a large extent the flux contributed by neighbouring objects is excluded by using segmentation maps and the  average flux within annuli is compensated for masked flux.  However, this process is less accurate for galaxies whose images are not perfectly axisymmetric. Figure \ref{fig:growth_curves} shows that the growth curves do not all level off perfectly at larger diameters due to random variations in the faint background.  Our method largely corrects for this by applying a mean correction to the magnitude measured using a slightly smaller aperture than that required to include virtually all the flux. It also has the additional advantage that it does not assume any particular surface brightness profile, \citep[e.g. a de Vaucouleurs profile for red galaxies as in ][]{brown07}.

As a check on our procedure, we compared our corrected apparent magnitudes with those obtained using apertures $\sim50\%$  larger in area than our preferred values and found that the systematic offset between the two was in general less than $\sim 0.05$ mag for $I>22.0$ and $\sim 0.02$ mag for $I<22.0$.  We thus concluded that the specific choice of aperture diameter does not greatly impact our results and conclusions.

\begin{figure}
 	\centering
		\includegraphics[width=0.4\textwidth]{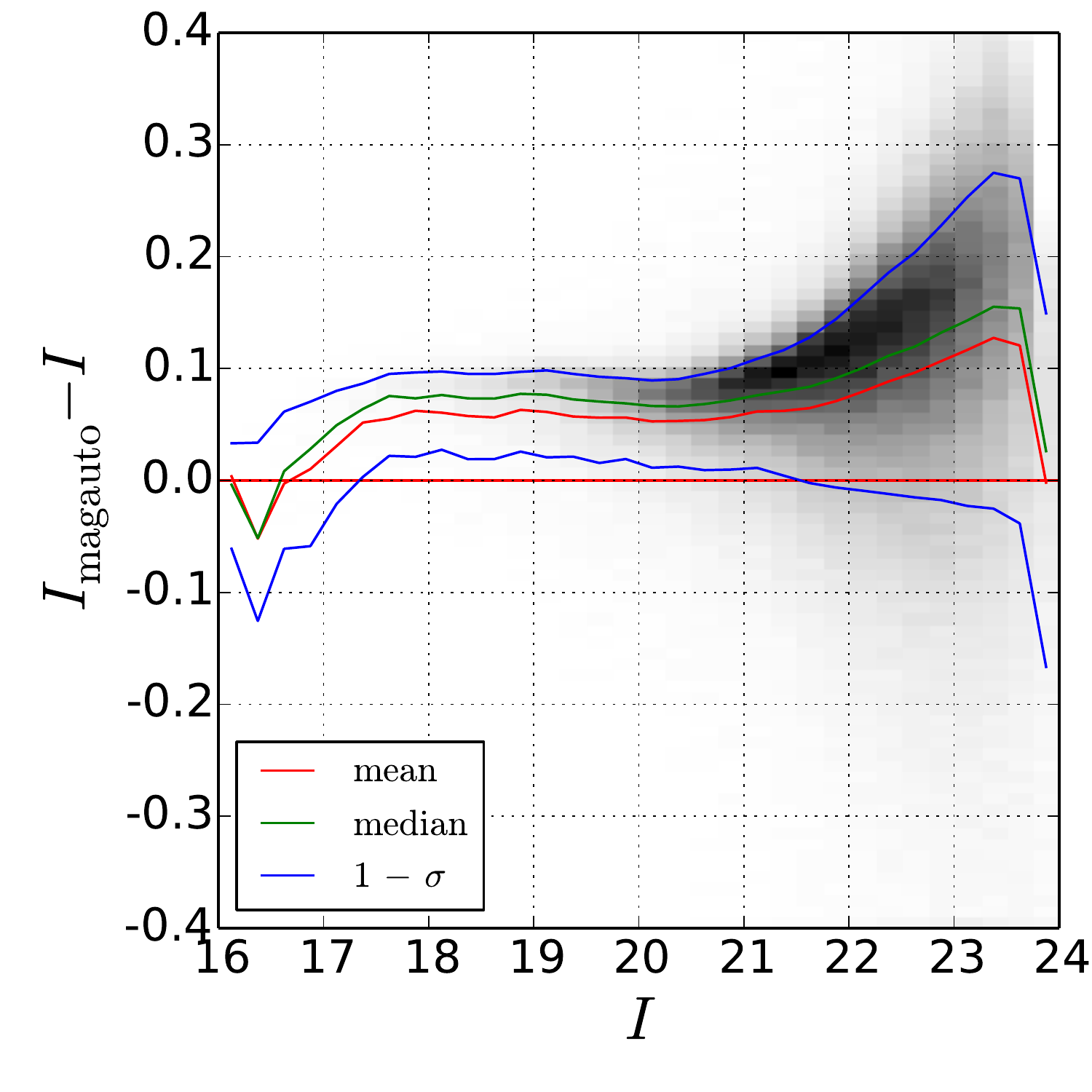}	
		\caption{Our apparent magnitudes are $0.06$ mag or more systematically brighter than those produced by MAG\_AUTO  because our method accounts for more of the total light from galaxies, especially faint ones for which MAG\_AUTO does not make a PSF correction.}
		\label{fig:magauto}
\end{figure}

As a second check we also compared our measured $I$-band magnitudes with those produced by SExtractor`s MAG\_AUTO and found that our values were systematically brighter by  $0.06$ mag or more as Figure \ref{fig:magauto} illustrates.  We attribute this to our improved estimates of the aperture required to capture the majority of the light together with improved corrections for any remaining missing light.  As first pointed out by \citet{labbe03}, the difference is particularly marked for galaxies fainter than $I \sim 20.5$  for which MAG\_AUTO does not make a PSF correction for light falling outside the photometric aperture. \citet{brown07} obtained a similar upturn at faint magnitudes in the difference between MAG\_AUTO magnitudes and 4 arcsecond aperture magnitudes for red galaxies (their Figure 1), but their offsets are \s0.05 mag smaller than ours.  We attribute this difference to our varying aperture size.

\begin{deluxetable}{cccc}
\tablewidth{0pt}
\tablecolumns{4}
\tabletypesize{\tiny}
\tablecaption {Apparent magnitude corrections including Moffat PSF correction.}
\tablehead{
\colhead{ $I$ } & \colhead{aperture diameter} & \colhead{aperture diameter} & \colhead{correction} \\
\colhead{ (4 arcsecond) } & \colhead{ to include most of light\tablenotemark{a}} & \colhead{ used} & \colhead{ applied} \\
\colhead{ (mag) } & \colhead{ (arcsecond) } & \colhead{ (arcsecond)} & \colhead{ (mag)}
 }
 \startdata
23.5 & 4 & 3 & -0.410 \\
22.5 & 5 & 4 & -0.243 \\
21.5 & 6 & 6 & -0.105 \\
20.5 & 8 & 8 & -0.070 \\
19.5 & 10 & 10 & -0.078 \\
18.5 & 15 & 15 & -0.061 \\
\enddata
\label{tab:apertures}
\tablenotetext{a}{\footnotesize{Diameter such that the magnitude as a function of diameter changes by less than 0.03  $\rm{\, mag \, arcsecond}^{-1}$.}}
\end{deluxetable}

\subsection{The templates}
\label{sec:templates}

We used the \citet{brown14} atlas of 129 ultraviolet to  mid-infrared spectral energy distributions (SEDs) of nearby galaxies for determining our photometric redshifts and for calculating the K-corrections used to determine absolute magnitudes. These templates combine ground-based and space-based observations in 26 photometric bands, gaps in spectral coverage being filled using MAGPHYS models \citep{cunha08}. The atlas spans a broad range of absolute magnitudes ($-14.7 < M_g < -23.2$) and colors ($0.1 < u-g < 1.9$). The systematic offsets and standard deviations for the residuals between the actual observed magnitudes and the observed magnitudes predicted by integrating each SED over the filter transmission curves are all less than 0.03 mag in the $ugriz$ wavebands except for the $u$-band standard deviation which is 0.06 mag. This provides a high degree of accuracy for our redshift and absolute magnitude calculations. The templates span the range of galaxy colors significantly better than previous SED libraries and we refer the reader to \citet{brown14} for a fuller discussion of how the template SEDs compare with observed photometry.

\vspace{20 pt}

\section{Photometric redshifts}
\label{sec:photoz}

Following \citet{brown07}, we originally intended to use photometric redshifts $z_\textrm{{phot}}$ determined using the empirical ANNz artificial neural network redshift code \citep{firth03, colli04}. However,  we found that the ANNz redshifts for fainter blue galaxies ($I \gtrsim 20.5$) exhibited a significant ($\sim 30\%$) deficiency in numbers at $z_{\rm{phot}} \sim 0.55$. This deficiency was also very clearly visible in any binned color-redshift plots that we made. We concluded that ANNz redshifts for blue galaxies  the range $0.5 \lesssim z_{\rm{phot}} \lesssim 0.6$ were either being shifted to greater or smaller values, or that ANNz was not producing valid redshifts for some blue galaxies, or a combination of both.   We therefore decided to use photometric redshifts determined using least-squares fits of the \citet{brown14} SEDs to our $uB_wRIyzJHK_s$ optical and infrared photometry and our IRAC infrared photometry in the 3.6, 4.5, 5.8 and 8.0 micron wavebands.

\begin{figure}
 	\centering
		\includegraphics[width=0.49\textwidth]{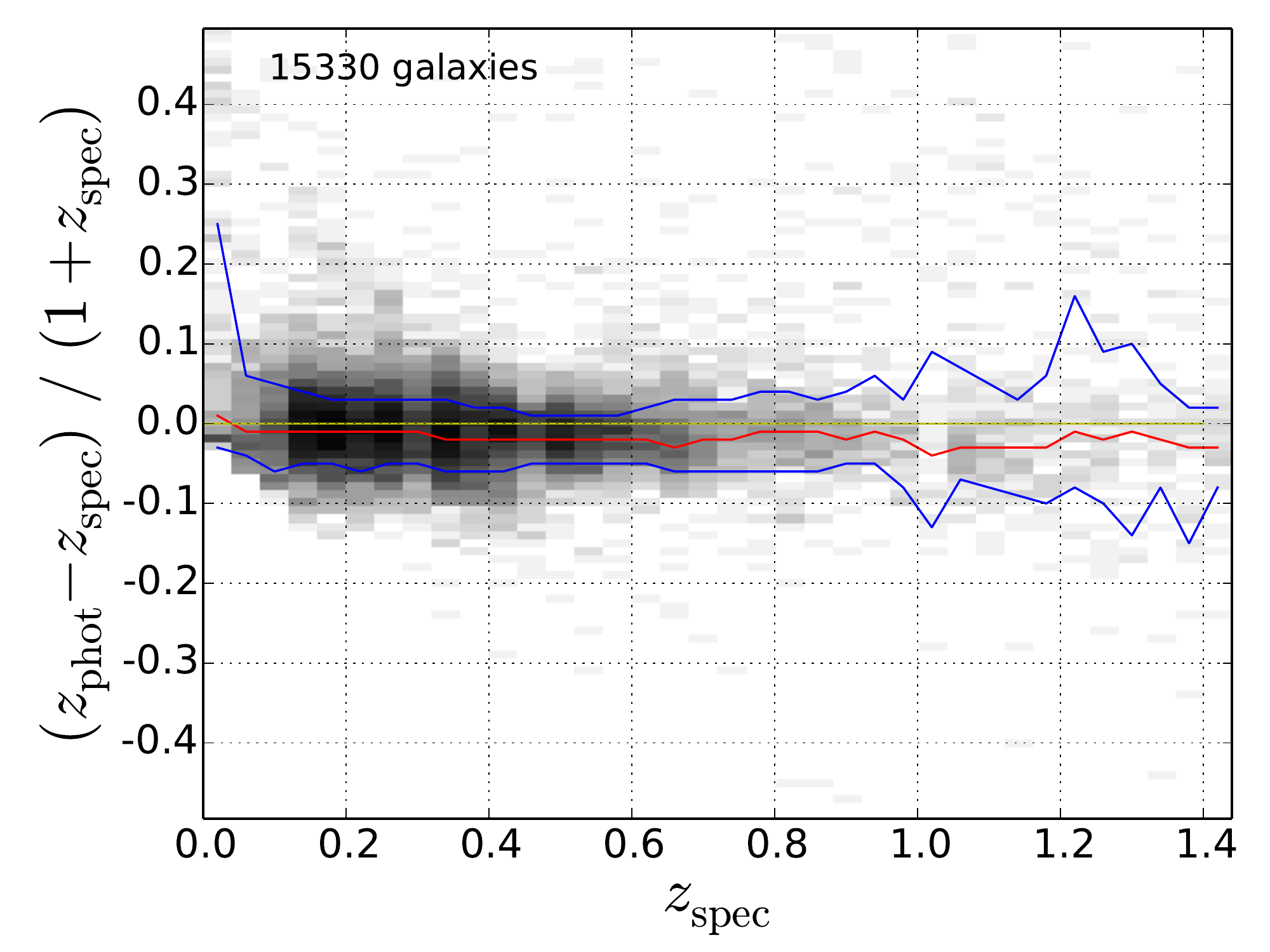}
		\caption{Systematic errors in $(z_{\rm{phot}}-z_{\rm{spec}}) / (1 + z_{\rm{spec}})$ are less than $\sim-0.03$ for $z < 1$. Random errors in $(z_{\rm{phot}}-z_{\rm{spec}}) / (1 + z_{\rm{spec}})$ are less than 0.05 for $0.2 \leq z < 0.9$ and up to $\sim0.1$ for $0.9 \leq z < 1.4$.  Red denotes the median error and blue the $1-\sigma$ deviations.}
		\label {fig:photoz_errors}
\end{figure}

Figure \ref{fig:photoz_errors} and Table \ref{tab:specznumbers} compare our photometric redshifts with available spectroscopic redshifts in \bootess (excluding AGN), and demonstrate that overall the systematic error in $(z_{\textrm{phot}}-z_{\textrm{spec}}) / (1+z_{\textrm{spec}})$ is less than -0.03 for $0.2 \leq z < 1.4$, while the random error in $(z_{\textrm{phot}}-z_{\textrm{spec}}) / (1+z_{\textrm{spec}})$ is 0.05 or less for $0.2 \leq z < 0.9$ and up to $\sim0.1$ for $0.9 \leq z < 1.4$.  We have included galaxies at redshifts $0 < z < 0.2$ and $1.2 \leq z < 1.4$ in Figure \ref{fig:photoz_errors} and Table \ref{tab:specznumbers} because galaxies at these redshifts can end up being counted in our range of interest ($0.2 \leq z < 1.2$) if their redshifts are significantly in error.

Catastrophic redshift failures were defined by $|z_{\textrm{phot}}-z_{\textrm{spec}}| > 0.15 (1+z_{\textrm{spec}})$ as in \citet{ilber13}. Table \ref{tab:specznumbers} indicates that the percentage of catastrophic redshift failures for red and blue galaxies together rises with redshift from $\sim1\%$ at $z\sim0.3$ to $\sim20\%$ at $z\sim1.3$. However, the numbers of galaxies involved at higher redshifts are small so that the corresponding percentages are significantly less certain. For the whole of our redshift range of interest ($0.2 \leq z < 1.2$) the percentage of catastrophic redshift failures is significantly lower for red galaxies than for blue, as one would expect from the tightness of the red sequence in color-color space. 

The accuracy of our photometric redshifts is improved by the fact that galaxies at $z < 1.2$ can be expected to differ relatively little from the sample of local galaxies on which the template SEDs in \citet{brown14} are based. We note however that comparisons of photometric with spectroscopic redshifts are subject to bias if the latter are not representative.   For example, if redshifts are compared for only the most luminous galaxies, these can be expected to have more accurate photometric redshifts because of their smaller photometric uncertainties, thus giving an unduly optimistic picture of overall photometric redshift accuracy.

\begin{deluxetable}{rrrrr}				
\tablewidth{0pt}				
\tablecolumns{5}				
\tablecaption {Photometric redshift errors, spectroscopic redshift numbers and the percentage of catastrophic redshift failures.}				
\tablehead{				
\colhead{$z_{\textrm{min}}$} & 	\colhead{$z_{\textrm{max}}$} & 	\colhead{$\dfrac{z_{\rm{phot}}-z_{\rm{spec}}}{1 + z_{\rm{spec}}}$} & 	\colhead{$N$} & 	\colhead{\%\,catastrophic} \\  
\colhead{} & 	\colhead{} & 	\colhead{} & 	\colhead{} & 	\colhead{redshift failures} \\ 
}				
 \startdata				
\\				
\multicolumn{5}{l}{All galaxies}\\				
\\				
0 .0& 	0.2 & 	$-0.01\pm 0.07$ & 	$3993$ & 	$3.7$ \\
0.2 & 	0.4 & 	$-0.01\pm 0.04$ & 	$5878$ & 	$1.1$ \\
0.4 & 	0.6 & 	$-0.02\pm 0.03$ & 	$3281$ & 	$1.6$ \\
0.6 & 	0.8 & 	$-0.02\pm 0.04$ & 	$1116$ & 	$4.4$ \\
0.8 & 	1.0 & 	$-0.01\pm 0.05$ & 	$504$ & 	$10.1$ \\
1.0 & 	1.2 & 	$-0.03\pm 0.08$ & 	$314$ & 	$15.0$ \\
1.2 & 	1.4 & 	$-0.02\pm 0.10$ & 	$201$ & 	$19.9$ \\
\\				
\multicolumn{5}{l}{Red galaxies}\\				
\\				
0.0 & 	0.2 & 	$0.01\pm 0.07$ & 	$1384$ & 	$4.9$ \\
0.2 & 	0.4 & 	$-0.01\pm 0.04$ & 	$2700$ & 	$0.6$ \\
0.4 & 	0.6 & 	$-0.02\pm 0.03$ & 	$1689$ & 	$0.4$ \\
0.6 & 	0.8 & 	$-0.02\pm 0.03$ & 	$568$ & 	$0.7$ \\
0.8 & 	1.0 & 	$-0.02\pm 0.03$ & 	$205$ & 	$3.9$ \\
1.0 & 	1.2 & 	$-0.03\pm 0.05$ & 	$117$ & 	$4.3$ \\
1.2 & 	1.4 & 	$-0.04\pm 0.11$ & 	$54$ & 	$5.6$ \\
\\				
\multicolumn{5}{l}{Blue galaxies}\\				
\\				
0 .0& 	0.2 & 	$-0.01\pm 0.06$ & 	$2609$ & 	$3.0$ \\
0.2 & 	0.4 & 	$-0.02\pm 0.04$ & 	$3178$ & 	$1.6$ \\
0.4 & 	0.6 & 	$-0.02\pm 0.04$ & 	$1592$ & 	$2.8$ \\
0.6 & 	0.8 & 	$-0.01\pm 0.05$ & 	$548$ & 	$8.2$ \\
0.8 & 	1.0 & 	$-0.01\pm 0.06$ & 	$299$ & 	$14.4$ \\
1.0 & 	1.2 & 	$-0.03\pm 0.10$ & 	$197$ & 	$21.3$ \\
1.2 & 	1.4 & 	$-0.02\pm 0.11$ & 	$147$ & 	$25.2$ \\
\\				
\enddata				
\label{tab:specznumbers}				
\end{deluxetable}				

\section{Sample selection}
\label{sec:selection}

As already noted in Section \ref{sec:apertures}, regions surrounding very extended galaxies and saturated stars were removed from our field, and this occurred before our initial sample was generated.  We then applied the cuts listed in Table \ref{tab:cuts} to restrict our sample to galaxies with good quality data.

\begin{deluxetable*}{llll}
\tablewidth{0pt}
\tablecolumns{3}
\tabletypesize{\scriptsize}
\tablecaption {Cuts used to select objects and separate red and blue galaxies.}
\tablehead{
\colhead{cut} & \colhead{purpose} & \colhead{cuts}  & \colhead{number (per cent)} \\
\colhead{number} & \colhead{of cut} & \colhead{} & \colhead{excluded} 
}
 \startdata
 \\
1 & exclude faint $I$-band objects & $I > 24.0 $ &  894\,962 (54.73\%) \\ [4pt]
2 & exclude objects with faint near infrared & $[3.6 \mu \rm{m}] > 23.3 $ &  181\,561 (11.10\%)  \\ [4pt]
3 & exclude stars & $ (R-I) > 0.683 + 0.5(I-[3.6 \mu \rm{m}]) $ & 56\,455 (3.45\%)    \\ [4pt]
4 & exclude stars  & $I_{\rm{2 arcsecond}} - I_{\rm{3 arcsecond}} \leq 0.4 $ &  2922 (0.18\%) \\[4pt]
5 & exclude AGN & $([3.6 \mu \rm{m}]-[4.5 \mu \rm{m}]) > 0.128$  & 11\,066 (0.68\%)  \\
&  (using modified & and $([5.8 \mu \rm{m}]-[8.0 \mu \rm{m}]) > -0.04$  &   \\ 
&  \citet{stern05} cuts) & and $([3.6 \mu \rm{m}]-[4.5 \mu \rm{m}]) > -2.272 + 2.5 ([5.8 \mu \rm{m}]-[8.0 \mu \rm{m}] - 0.96)$  &   \\ [4pt]
6 &  \multicolumn {2}{l}{exclude further AGN identified by SDSS and AGES} & 124 (0.01\%)  \\ [4pt]
7 & restrict abs mag range &  $ -25.0 \leq M_{\rm{B}} < -15.0 $  &  79\,794 (4.88\%)  \\ [4pt]
8 & red-blue separation  &  $ (M_{\rm{U}} - M_{\rm{B}}) > 1.074 - 0.18z  - 0.03(M_{\rm{B}} + 19.62)$   & 0 (0.00\%)  \\ [4pt]
\hline\\
 &  number remaining  & & 408\,495 (24.98\%)\\
\enddata
\label{tab:cuts}
\end{deluxetable*}

\subsection{Apparent magnitude limits}
\label{sec:appmag_limits}

Our primary magnitude cuts are $I < 24.0$ and [3.6 \muu m] $ < 23.3$, which provide us with a highly complete sample with reliable photometric redshifts (Section \ref{sec:photoz}).  The [3.6 \muu m] cut excludes objects for which the [3.6 \muu m] uncertainties would be large as this waveband is important for the accuracy of our photometric redshifts. All objects in our sample have good NDWFS, NEWFIRM and IRAC imaging. We correct for $I$-band incompleteness using the method in \citet{brown07} which is described below (Section \ref{sec:completeness}). Once we have applied the $I < 24.0$ limit, we find that only 0.8\% of our sources have [3.6 \muu m] > 23.3 and therefore do not apply a correction for  incompleteness in the 3.6 \muu m band.

\begin{figure}
 	\centering
		\includegraphics[width=0.45\textwidth]{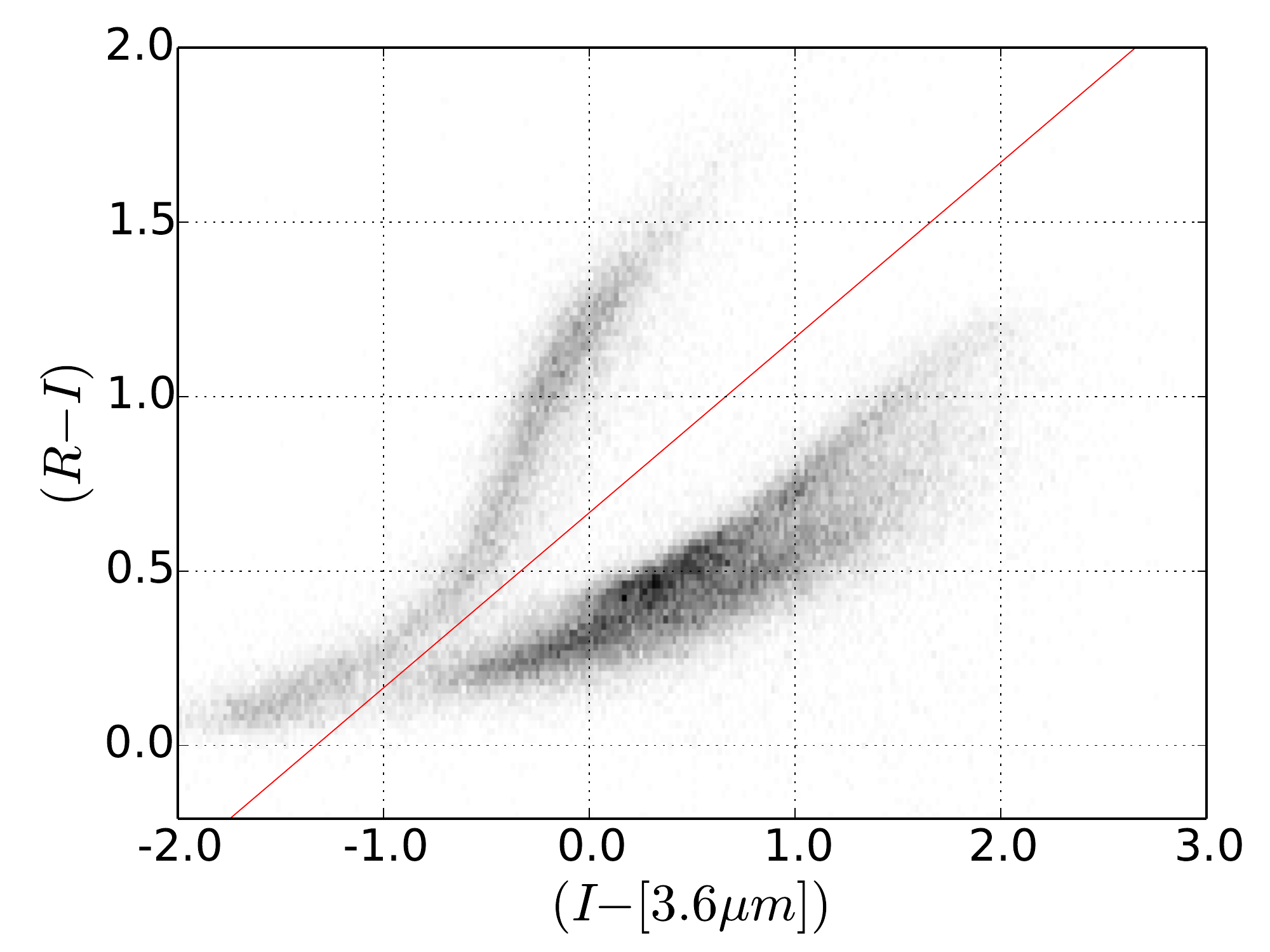}	
		\caption{Excluding stars. The color-color cut $(R - I) \geq 0.683 + 0.5(I - [3.6 \mu m])$ effectively removes stars from the sample, except for a small amount of overlap with very blue galaxies with $(I - [3.6 \mu \rm{m}]) \leq -0.5$.}
		\label{fig:star_cuts}
\end{figure}

Because stellar properties are well defined, stars form a tight sequence in $(R - I)$ versus $(I - [3.6 \mu m])$ color-color space and could therefore be excluded using the simple cut shown in Figure \ref{fig:star_cuts}. As confirmation of the effectiveness of this cut in removing stars, we additionally plotted the difference in measured $I$-band magnitudes when using apertures of diameter 2 and 3 arcsecond.  Objects classified as stars by our color cut appeared as a thin horizontal locus of constant magnitude difference identical to that to be expected for point sources with our chosen point spread function, confirming that they are indeed stars (or possibly quasars, but we exclude these with a further cut). Although the numbers involved were small, we additionally removed objects for which the difference in $I$-band magnitudes was more than 0.4 mag for the 2 and 3 arcsecond diameter measurement apertures.

\begin{figure*}
 	\centering
		\includegraphics[width = 0.9\textwidth]{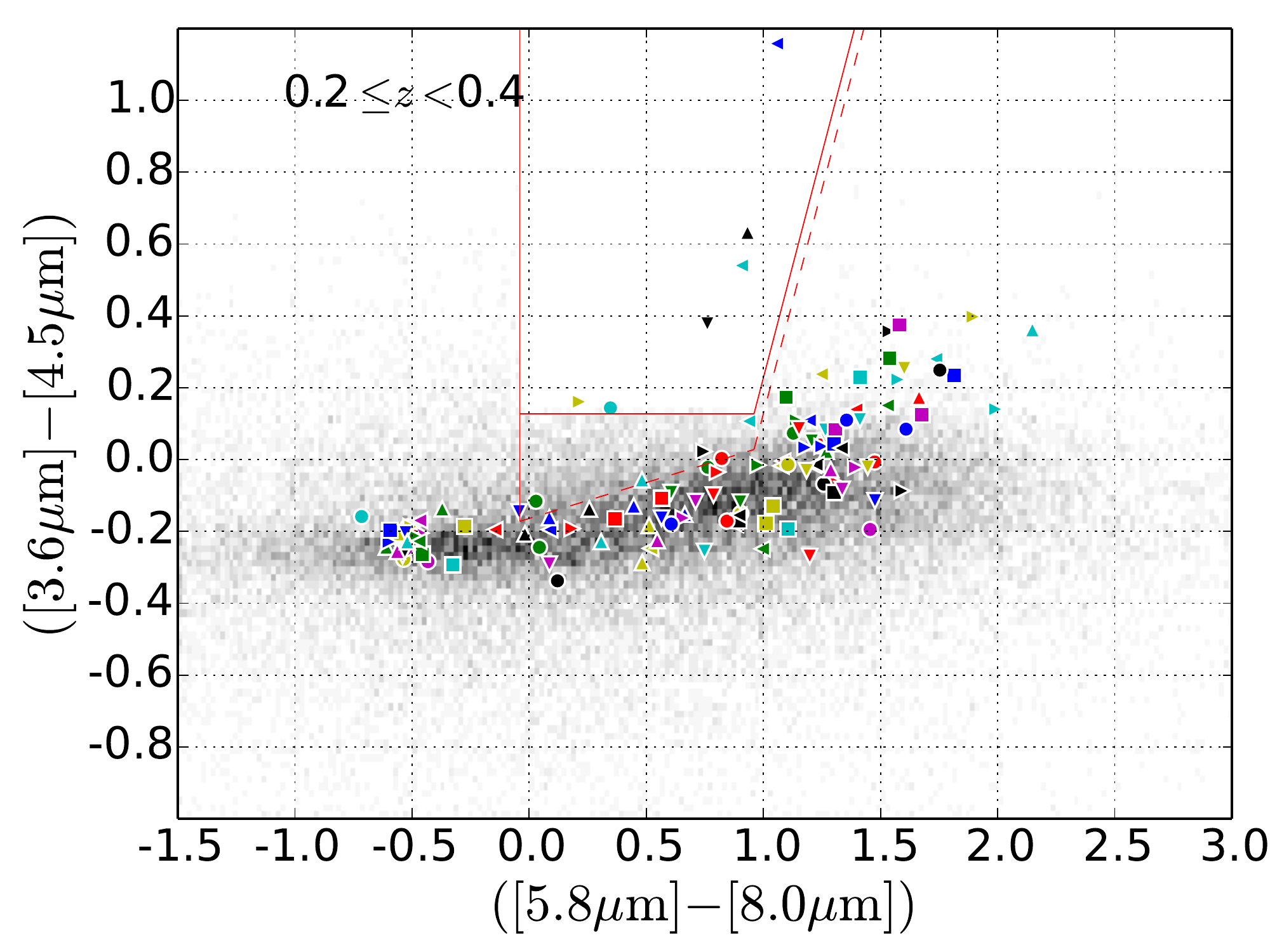}
		\caption{Example plot for $0.2 \leq z_{\textrm{phot}} < 0.4$ showing how AGN are excluded. Our three IRAC color-color cuts (solid lines) are modifications of those used by \citet{stern05} to select \textit{for} AGN rather than exclude them (dashed lines).  Colored markers indicate our template galaxies at $z_{\textrm{phot}} = 0.3$. The galaxies appearing significantly redder in ([5.8 $\mu$m] - [8.0 $\mu$m]) color than the template locus have artificially high red colors due to noise in the 8.0 $\mu$m band.}
\label{fig:AGN}
\end{figure*}

Type I and Type II AGN were excluded by the three cuts in $([3.6 \mu m] - [4.5 \mu m])$ versus $([5.8 \mu m] - [8.0 \mu m])$ color-color space shown in Figure \ref{fig:AGN} and Table \ref{tab:cuts}.  These cuts are similar to those used by \citet{stern05} to select \textit{for} AGN, rather than to exclude them as we do; however, we have raised their middle cut by 0.2 mag to prevent it from removing significant numbers of galaxies which do not have AGN.  Our cuts do result in a small number of AGN not being excluded that should be, and for this reason any galaxies classified as AGN by AGES or SDSS are also excluded. Figure \ref{fig:AGN} also shows that only a few of our template galaxies would be excluded by our cuts and these are ones known to contain AGN. We see from Table \ref{tab:cuts} that the fraction of galaxies classified as AGN is no more than \s2\% so that further refining our classification of AGN would not significantly impact our results.

\subsection{Separating red and blue galaxies}
\label{sec:redbluecut}

\begin{figure*}
 	\centering
		\includegraphics[width=0.9\textwidth]{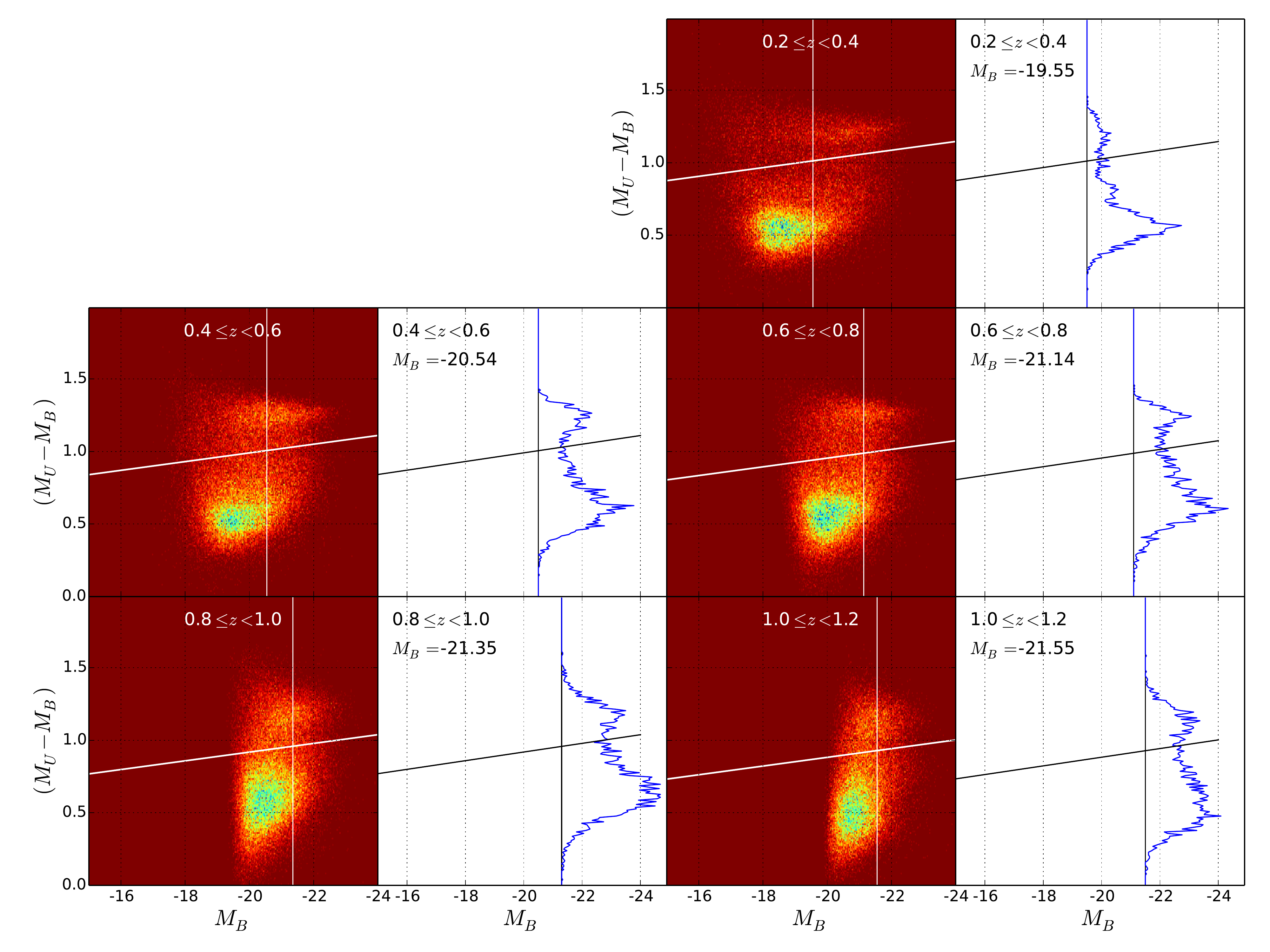}	
		\caption{How an evolving cut in the $(M_{\rm{U}}-M_{\rm{B}})$ versus $M_{\rm{B}}$ plane is used to separate red and blue galaxies (Equation \ref{eq:redbluecut}). The histograms on the right plot the relative numbers of galaxies of different restrame colors at the absolute magnitudes indicated by the vertical white lines and show graphically how the position of the green valley was determined. They indicate that bimodality is present even in the highest redshift bin.}
\label{fig:redblue}
\end{figure*}

As Figure \ref{fig:redblue} shows, bimodality is evident beyond $z = 1$ on $(M_{\rm{U}}-M_{\rm{B}})$ versus $M_{\rm{B}}$ color-magnitude plots.  We separated red and blue galaxies using an evolving empirical cut through the centre of the green valley, the position of which was determined for each redshift bin from histograms of the relative numbers of galaxies with different restrame colors at a fixed absolute magnitude. In each redshift bin this fixed magnitude was chosen so as to intersect both the red sequence and the blue cloud and is indicated by a vertical white line in Figure \ref{fig:redblue}. This resulted in our definition of a red galaxy as one for which:

\begin{equation}\label{eq:redbluecut}
	(M_{\rm{U}} - M_{\rm{B}}) > 1.074 - 0.18z  - 0.03(M_{\rm{B}} + 19.4). \\
\end{equation}
            
In their determinations of $B$-band luminosity functions, \citet{willm06} and \citet{faber07} used a similar but redshift independent $(M_{\rm{U}}-M_{\rm{B}})$ versus $M_{\rm{B}}$ cut, which was approximately 0.05 mag above our own: $(M_{\rm{U}} - M_{\rm{B}}) > 0.419 - 0.032 (M_{\rm{B}} + 21.52) - 0.52$. \citet{bell04} used a redshift-dependent red-blue cut based on a plot of $(M_{\rm{U}}-M_V)$ versus $M_V$, and \citet{brown07} used a very similar cut, but we prefer to use $(M_{\rm{U}}-M_{\rm{B}})$ versus $M_{\rm{B}}$ because it gives clearer bimodality with our data set.  
 
We checked the dependence of our measured Schechter luminosity function parameters  (Section \ref{sec:lumfn}) on the exact position of the red-blue cut. We found that varying the cut up or down by 0.05 mag made less than 16\% difference to the space density parameter $\phi^*$,  less than 16\% difference to the luminosity density $j_{\rm{B}}$ of red galaxies, and less than 6\% difference to that of blue galaxies. For the characteristic magnitude parameter $M^*$ and the measured magnitude (Section \ref{sec:results_luminous}) of the very brightest galaxies the variations were no more than 0.06 mag, and generally much less, especially for red galaxies.  

\subsection{Completeness correction}
\label{sec:completeness}

At the apparent magnitudes of interest the completeness is largely determined by source confusion rather than objects being lost in background noise.  \citet{brown07} measured the completeness for red galaxies in the \bootess field by adding mock galaxies to their catalogue and then attempting to recover them.  They found that the completeness, as a function of observed $I$-magnitude, was well described by $C_I(I) = 1 - 0.05(I-21.5)$ for $21.5 < I \leq 24.0$ and $C_I(I) = 1$ for $I \leq 21.5$.  We assume that the same formula applies when blue galaxies are included.  Our apparent magnitude limit of $I = 24.0$ [for which $C_I(24.0) = 0.875$] is designed to ensure completeness of better than 85\%. As already indicated in \S\ref{sec:appmag_limits}, the 3.6 \muu m band completeness is over 99\% once the $I<24.0$ cut has been applied so we do not attempt to estimate its exact value.

By considering the numbers of galaxies  $n(M_B, I) \Delta M_B \Delta I$ in absolute magnitude-apparent magnitude bins and summing over apparent magnitudes we find that completeness as a function of absolute magnitude $M_B$ varies between 88\% and 100\% as given by the following formula:

\begin{equation} \label{eq:totalcompleteness}
C_M(M_B) \; = \;
	{\displaystyle\sum_{I}^{} n(M_B, I) }
	\; \Big/ \;
	{\displaystyle\sum_{I}^{} n(M_B, I) / C_I(I) }.
\end{equation}

\section{Calculation of absolute $U$ and $B$ magnitudes}
\label{sec:absmags}

\begin{figure}
 	\centering
		\includegraphics[width=0.45\textwidth]{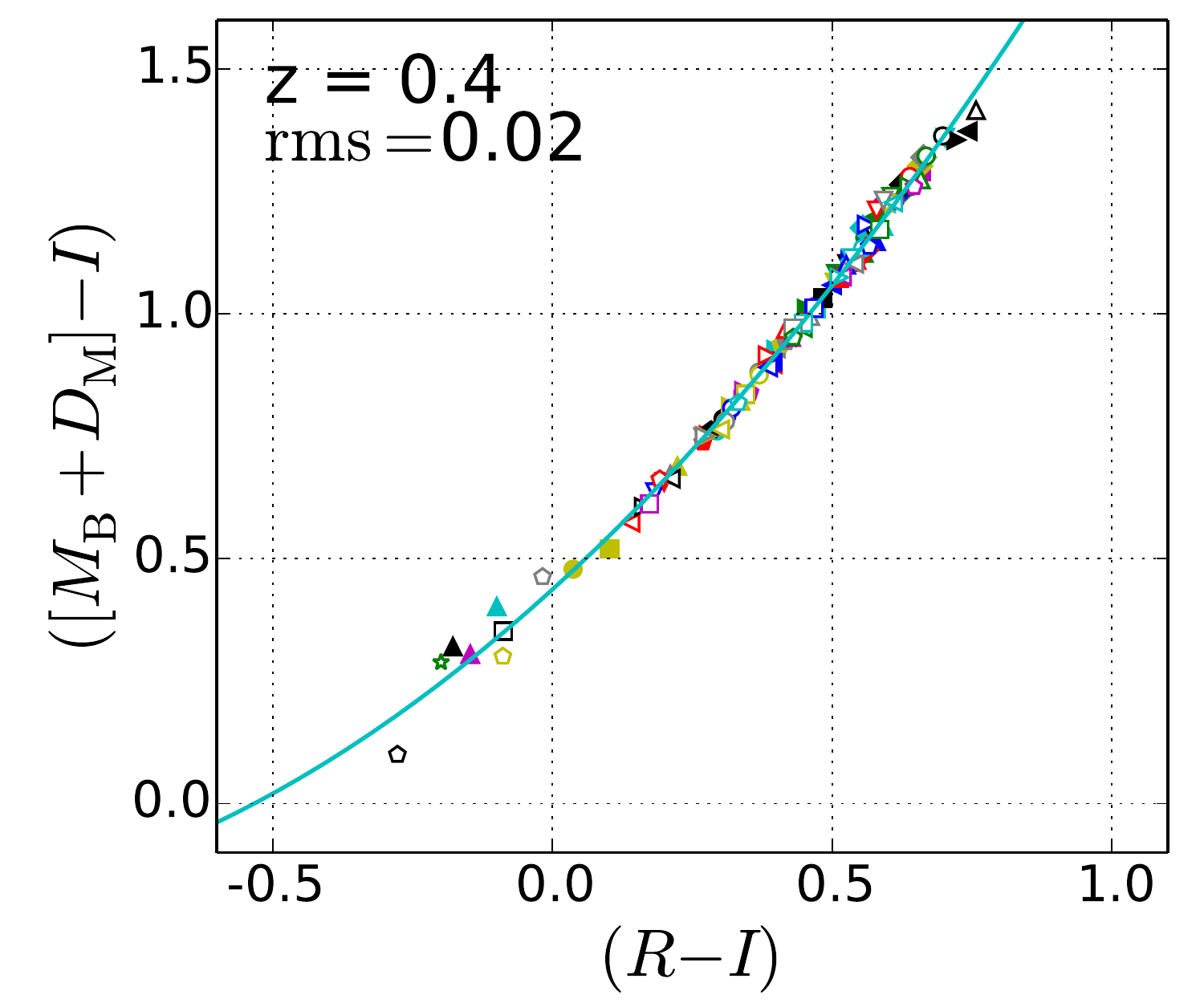}
		\caption{Example plot showing how we determine absolute magnitudes from observed colors using the method of \citet{beare14}. The colored markers plot computed values of ${K_{BI} \equiv (M_B + D_{\rm{M}}) -  I}$  against ${(R - I)}$ for the 129 template SEDs from \citet{brown14} at $z=0.4$. $ D_{\rm{M}}$ is the distance modulus.  The curve is the best fit second degree polynomial to the template data points and enables absolute magnitudes $M_B$ to be determined from apparent $R$ and $I$ magnitudes. The RMS offset from the template points is shown in the top left corner. Outliers offset by more than 0.2 mag from the polynomials are excluded from the polynomial fitting.}
		\label{fig:calibration_example}
\end{figure}

We used the method of \citet{beare14}, calculating the absolute magnitude $M_W$ in a waveband $W$ from second degree polynomial fits at different redshifts to plots of $(M_W + D_M) - m_Z$ against a carefully chosen observed color $(m_Y - m_Z)$, with $D_M$ being the distance modulus. Figure \ref{fig:calibration_example} shows an example plot and Table \ref{tab:observed_colors} lists the observed colors we used at different redshifts to determine $M_U$ and  $M_B$. We make the polynomial coefficients used to calculate $U$ and $B$-band (and also $V$ and $g$-band) absolute magnitudes from observed colors available in full on-line\footnote{https://dx.doi.org/10.4225/03/563930353DA9E} together with the corresponding polynomial plots like that in Figure \ref{fig:calibration_example}. The RMS scatter of the templates about the fits is less than 0.05 for both $M_U$ and $M_B$ across the entire redshift range.

\begin{deluxetable}{cccc}	
\tablewidth{0pt}
\tablecolumns{4}
\tabletypesize{\scriptsize}
\tablecaption {The observed colors used to determine absolute $U$ and $B$ magnitudes.}
\tablehead{
\colhead{restframe} & \colhead{redshift} & \colhead{color} & max{}
\\
\colhead{waveband} & \colhead{range} & \colhead{$(m_Y - m_Z)$} & \colhead{RMS}
 \\
\colhead{$M_W$} & \colhead{} & \colhead{} & \colhead{offset}
}
 \startdata
$U$ & 0.0 to 0.8 & $(B_w - R)$ & 0.049\\
$U$ & 0.8 to 1.2 & $(R - I)$ & 0.026\\
\\
$B$ & 0.0 to 0.4 & $(B_w - R)$ & 0.040\\
$B$  & 0.4 to 0.8 & $(R - I)$ & 0.023\\
$B$ & 0.8 to 1.2 & $(I - J)$ & 0.037\\
\enddata
\label{tab:observed_colors}
\tablecomments{Absolute magnitudes in a waveband $W$ are calculated using the method of \citet{beare14}. Given two suitably chosen observed magnitudes $m_Y$ and $m_Z$, $(M_W - m_Z)$ is given by a second degree polynomial in the color $(m_Y - m_Z)$. The polynomial coefficients are available on-line.}
\end{deluxetable}

\vspace{20 pt}
									
\section{Determination of $B$-band luminosity functions}
\label{sec:lumfn}

LFs were determined for red and blue galaxy sub-samples separately as well as for the total sample. In each case, galaxies in the redshift range $0.2 \leq z < 1.2$ were allocated to five redshift bins of equal width $\Delta z = 0.2$. For each redshift bin, empirical binned LFs, $\Phi(M)$,  were obtained by dividing the (completeness corrected) numbers of galaxies $N$ in $B$-band absolute magnitude bins of width $\Delta M$ by the comoving volume $\Delta V$ corresponding to the given redshift range $\Delta z$, i.e.:

\begin{equation}
	\Phi(M) = N / \Delta V.
\label{eq:Phi}
\end{equation}

Parameters $\phi^*, M^*, \alpha$ for the best fitting Schechter function $\phi(M)$ to the binned luminosity function $\Phi(M)$ were obtained using a least squares $\chi^2$ fit minimisation technique. At the faint end we restricted ourselves to magnitudes $M < M_{\rm{faint}}$ for which at least 95\% of observed galaxies (when those with $I > 24.0$ are also included) had apparent magnitudes brighter than our faint limit of $I = 24.0$. $M_{\rm{faint}}$ was determined from plots of $M_B + \rm{D}_{\rm{M}} - I$ against redshift. We also confirmed that our $[3.6 \mu \rm{m}] < 23.3$ cut did not eat into the sample.  No bright end limit was used. Because our sample was at least 95\% complete at all apparent $I$-band magnitudes within each redshift bin, we did not need to use the $1/V_{\rm{max}}$ method to correct for varying completeness, but could use a simple best fit in each redshift bin to the numbers in different absolute magnitude bins in order to provide a first estimate of the best fit Schechter function.

These least squares best-fit Schechter parameters were used as starting points for determining maximum likelihood Schechter fits \citep[e.g.][]{marsh83} to the magnitude-redshift distribution $(M, z)$ within each redshift bin.  The $I$-band completeness correction described in Section \ref{sec:completeness} was included in the analysis.  We used the same faint limit $M_{\rm{faint}}$ as for the least squares fits. No bright end limit was used.

The space densities of faint galaxies become increasingly hard to determine accurately at higher redshifts where sample completeness drops rapidly and apparent magnitudes and photometric redshifts become increasingly uncertain.  For this reason, $\alpha$, which determines the faint end slope of the Schechter function, becomes increasingly hard to measure as redshift increases.  As we are unable to measure any possible evolution of  $\alpha$, we adopted fixed $\alpha$ values of $-0.5, -1.3$ and $-1.1$ for red, blue and all galaxies respectively, based on their best fitting maximum likelihood values in the two lowest redshift bins ($0.2 \leq z < 0.4$ and $0.4 \leq z < 0.6$). The values of $\phi^*$, $M^*$ and $\alpha$ when $\alpha$ is treated as a free parameter are presented in Table \ref{tab:results_variable}. 
		
\begin{deluxetable*}{rrrrr}				
\tablewidth{0pt}				
\tablecolumns{5}				
\tabletypesize{\tiny}				
\tablecaption {Maximum likelihood values of parameters, luminosity density and absolute magnitude of highly luminous galaxies when $\alpha$ is allowed to vary.}				
\tablehead{				
\colhead{z} & 	\colhead{$\alpha$} & 	\colhead{$\phi^*$} & 	\colhead{$M_B^* - 5\log h_{70}$} & 	\colhead{$M_B\left(10^{-4.0}\right) - 5\log h_{70}$}\\ 
\colhead{} & 	\colhead{} & 	\colhead{$/ \, 10^{-3} \, h_{70}^3 \, \rm{Mpc}^{-3} \, {\rm{mag}}^{-1}$} & 	\colhead{} & 	\colhead{}\\ 
\colhead{} & 	\colhead{} & 	\colhead{} & 	\colhead{} & 	\colhead{}\\ 
}				
 \startdata				
\\				
\multicolumn{5}{l}{All galaxies - variable $\alpha$}\\				
\\				
0.3 & 	$-1.24\pm 0.09$ & 	$4.40\pm 0.31$ & 	$-20.90\pm 0.07$ & 	$-22.24\pm 0.03$ \\
0.5 & 	$-1.10\pm 0.05$ & 	$5.50\pm 0.22$ & 	$-21.00\pm 0.09$ & 	$-22.44\pm 0.07$ \\
0.7 & 	$-1.18\pm 0.05$ & 	$4.71\pm 0.29$ & 	$-21.16\pm 0.08$ & 	$-22.53\pm 0.03$ \\
0.9 & 	$-1.59\pm 0.10$ & 	$4.14\pm 0.37$ & 	$-21.50\pm 0.10$ & 	$-22.69\pm 0.04$ \\
1.1 & 	$-1.81\pm 0.01$ & 	$2.59\pm 0.16$ & 	$-21.72\pm 0.05$ & 	$-22.69\pm 0.04$ \\
\\				
\multicolumn{5}{l}{Red galaxies - variable $\alpha$}\\				
\\				
0.3 & 	$-0.58\pm 0.20$ & 	$2.51\pm 0.13$ & 	$-20.63\pm 0.06$ & 	$-22.05\pm 0.03$ \\
0.5 & 	$-0.57\pm 0.15$ & 	$2.68\pm 0.16$ & 	$-20.81\pm 0.16$ & 	$-22.26\pm 0.06$ \\
0.7 & 	$-0.60\pm 0.17$ & 	$1.79\pm 0.19$ & 	$-21.01\pm 0.15$ & 	$-22.29\pm 0.03$ \\
0.9 & 	$-0.93\pm 0.09$ & 	$1.96\pm 0.19$ & 	$-21.26\pm 0.07$ & 	$-22.44\pm 0.04$ \\
1.1 & 	$-1.44\pm 0.07$ & 	$1.34\pm 0.13$ & 	$-21.51\pm 0.07$ & 	$-22.35\pm 0.03$ \\
\\				
\multicolumn{5}{l}{Blue galaxies - variable $\alpha$}\\				
\\				
0.3 & 	$-1.48\pm 0.19$ & 	$2.39\pm 0.23$ & 	$-20.90\pm 0.56$ & 	$-21.95\pm 2.15$ \\
0.5 & 	$-1.32\pm 0.04$ & 	$3.15\pm 0.24$ & 	$-20.99\pm 0.08$ & 	$-22.19\pm 0.08$ \\
0.7 & 	$-1.42\pm 0.06$ & 	$2.97\pm 0.21$ & 	$-21.19\pm 0.12$ & 	$-22.33\pm 0.03$ \\
0.9 & 	$-1.93\pm 0.12$ & 	$2.04\pm 0.28$ & 	$-21.64\pm 0.15$ & 	$-22.49\pm 0.04$ \\
1.1 & 	$-2.18\pm 0.07$ & 	$0.99\pm 0.19$ & 	$-21.97\pm 0.11$ & 	$-22.25\pm 0.06$ \\
\enddata				
\label{tab:results_variable}				
\end{deluxetable*}							

\begin{deluxetable*}{rrrrrr}					
\tablewidth{0pt}					
\tablecolumns{6}					
\tabletypesize{\tiny}					
\tablecaption {Maximum likelihood values of parameters, luminosity density and absolute magnitude of highly luminous galaxies when $\alpha$ is set to its values at $0.2 < z < 0.6$.}					
\tablehead{					
\colhead{z} & 	\colhead{$\alpha$} & 	\colhead{$\phi^*$} & 	\colhead{$M_B^* - 5\log h_{70}$} & 	\colhead{$M_B\left(10^{-4.0}\right) - 5\log h_{70}$} & 	\colhead{$j_B$} \\
\colhead{} & 	\colhead{} & 	\colhead{$/ \, 10^{-3} \, h_{70}^3 \, \rm{Mpc}^{-3} \, {\rm{mag}}^{-1}$} & 	\colhead{} & 	\colhead{} & 	\colhead{$/ \, 10^8 \, h_{70} \, L_{\sun} \, {\rm{Mpc}}^{-3}$} \\
\colhead{} & 	\colhead{} & 	\colhead{} & 	\colhead{} & 	\colhead{} & 	\colhead{}
}					
 \startdata					
\\					
\multicolumn{6}{l}{All galaxies - fixed $\alpha$}\\					
\\					
0.3 & 	$-1.1$  & 	$5.51\pm 0.25$ & 	$-20.74\pm 0.06$ & 	$-22.18\pm 0.04$ & 	$1.61\pm 0.09$ \\
0.5 & 	$-1.1$  & 	$5.50\pm 0.23$ & 	$-21.00\pm 0.08$ & 	$-22.44\pm 0.07$ & 	$2.05\pm 0.15$ \\
0.7 & 	$-1.1$  & 	$5.16\pm 0.09$ & 	$-21.09\pm 0.04$ & 	$-22.52\pm 0.04$ & 	$2.10\pm 0.07$ \\
0.9 & 	$-1.1$  & 	$6.05\pm 0.22$ & 	$-21.19\pm 0.05$ & 	$-22.67\pm 0.04$ & 	$2.70\pm 0.08$ \\
1.1 & 	$-1.1$  & 	$3.83\pm 0.19$ & 	$-21.35\pm 0.06$ & 	$-22.69\pm 0.05$ & 	$1.98\pm 0.08$ \\
\\					
\multicolumn{6}{l}{Red galaxies - fixed $\alpha$}\\					
\\					
0.3 & 	$-0.5$  & 	$2.67\pm 0.10$ & 	$-20.56\pm 0.04$ & 	$-22.03\pm 0.03$ & 	$0.55\pm 0.02$ \\
0.5 & 	$-0.5$  & 	$2.85\pm 0.10$ & 	$-20.74\pm 0.08$ & 	$-22.23\pm 0.06$ & 	$0.70\pm 0.03$ \\
0.7 & 	$-0.5$  & 	$1.89\pm 0.10$ & 	$-20.93\pm 0.05$ & 	$-22.29\pm 0.03$ & 	$0.55\pm 0.01$ \\
0.9 & 	$-0.5$  & 	$2.21\pm 0.20$ & 	$-21.02\pm 0.05$ & 	$-22.43\pm 0.04$ & 	$0.70\pm 0.04$ \\
1.1 & 	$-0.5$  & 	$1.52\pm 0.13$ & 	$-21.11\pm 0.05$ & 	$-22.38\pm 0.03$ & 	$0.52\pm 0.02$ \\
\\					
\multicolumn{6}{l}{Blue galaxies - fixed $\alpha$}\\					
\\					
0.3 & 	$-1.3$  & 	$3.53\pm 0.06$ & 	$-20.64\pm 0.05$ & 	$-21.88\pm 0.05$ & 	$1.15\pm 0.05$ \\
0.5 & 	$-1.3$  & 	$3.23\pm 0.16$ & 	$-20.97\pm 0.09$ & 	$-22.19\pm 0.08$ & 	$1.43\pm 0.14$ \\
0.7 & 	$-1.3$  & 	$3.46\pm 0.02$ & 	$-21.08\pm 0.05$ & 	$-22.31\pm 0.06$ & 	$1.69\pm 0.08$ \\
0.9 & 	$-1.3$  & 	$3.78\pm 0.11$ & 	$-21.22\pm 0.06$ & 	$-22.49\pm 0.04$ & 	$2.11\pm 0.05$ \\
1.1 & 	$-1.3$  & 	$2.50\pm 0.14$ & 	$-21.39\pm 0.09$ & 	$-22.52\pm 0.07$ & 	$1.63\pm 0.06$ \\
\enddata					
\label{tab:results_fixed}					
\end{deluxetable*}

\subsection{Sources of error } 
\label{sec:errors}

As with all large volume surveys, the largest source of error is cosmic variance \citep[e.g.][]{somer04, brown07}.  To estimate its effect on our measurements we chose nine subfields, each 0.7 deg square, or 16.9 times smaller than our total field area,   repeated our determinations of luminosity function evolution for each, and measured the standard deviations of our parameters.  We chose non-contiguous subfields in order to minimise  correlation between subfields due to structures such as cluster, filaments and voids overlapping two subfields.  Assuming no such correlation we would expect the numbers of galaxies in given redshift and absolute magnitude bins to be Poisson variables and the standard deviation for the whole field to be $\sqrt{16.9}$ times smaller than that between the individual subfields.  The cosmic variance in each redshift bin is $\sim3\%$. (The smallest and largest values are 1.8\% for $0.2 \leq z < 0.4$ and 4.3\% for $0.4 \leq z < 0.6$.) Using mock catalogues and the same photometry \citet{brown08} obtained cosmic variances for $0.2 \leq z < 1.0$ red galaxies of $\sim8\%$ within each redshift bin. They note that uncertainties derived from mock catalogs are typically $50\%$ larger and should be more robust than those obtained from subsamples because large-scale structures can span more than one subsample.  As one would expect from the fact that red galaxies are more strongly concentrated in clusters than blue, the cosmic variance for red galaxies in different redshift bins is up to twice as great as for all galaxies.

We determined cosmic variance errors for our Schechter parameters, and for luminosity density and the magnitude of the brightest galaxies, by fitting maximum likelihood Schechter functions for each of the nine subfields individually. As we see later in Section \ref{sec:results} cosmic variance errors do not significantly affect our results. However, the true errors are likely to be slightly larger as clustering features such as filaments may extend across more than one of our subfields, even though they are smaller than the whole \bootess field.

We also investigated the effect of the random photometric redshift errors shown in Figure \ref{fig:photoz_errors} and Table \ref{tab:specznumbers} on our maximum likelihood luminosity functions. We did this by convolving Gaussian functions representing the random photometric redshift errors (typically $\sigma_z \sim 0.03$) with our measured Schechter functions, and found that the change in magnitude at any fixed space density was less than 0.01 mag, except for the bright end of the luminosity function in the lowest redshift bin ($0.2 \leq z < 0.4$). However, most luminous galaxies in this redshift range have spectroscopic redshifts, and we use these in preference to photometric ones when available, so our errors should remain no more than \s0.01 mag at all redshifts.

\begin{figure}
 	\centering
	\includegraphics[width=0.45\textwidth]{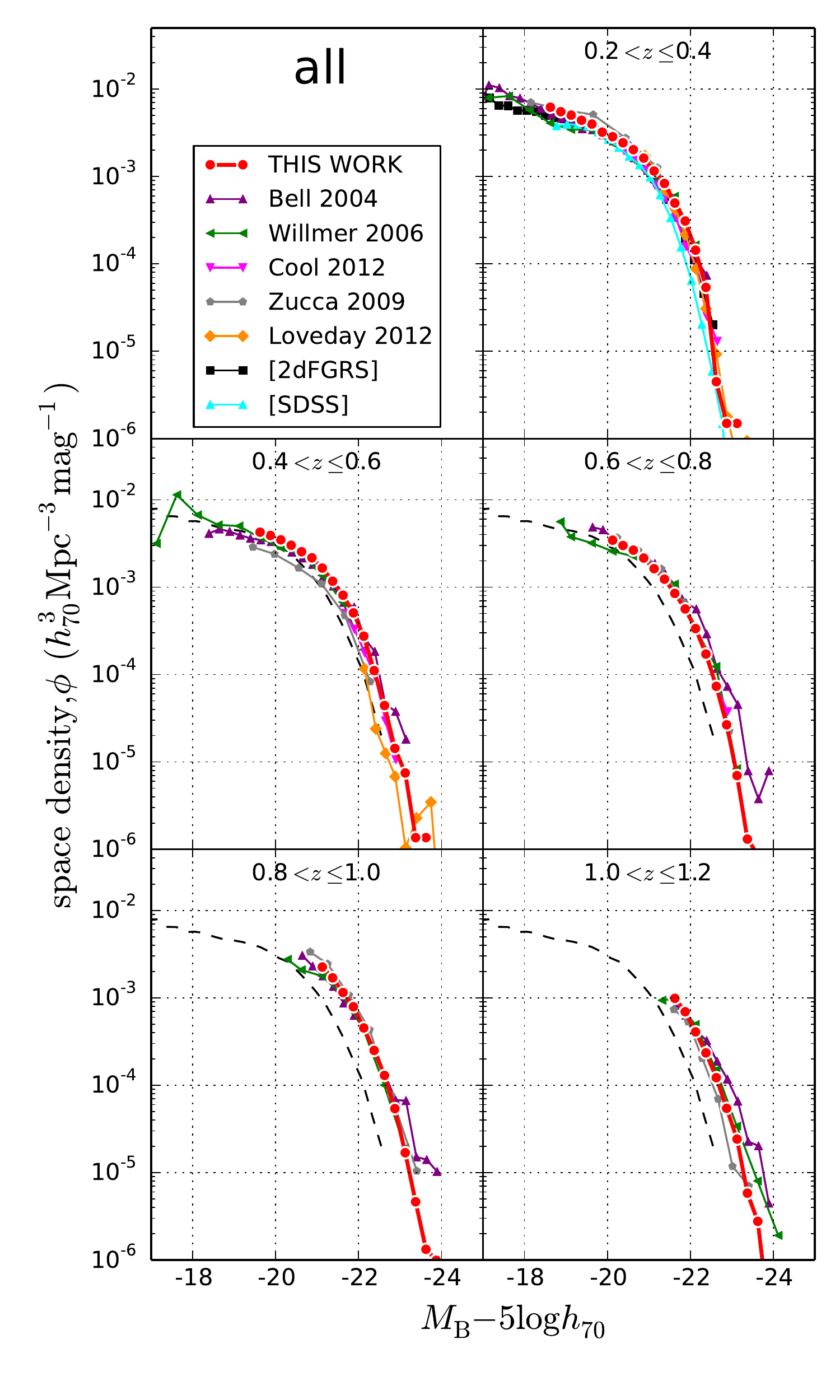}
	\caption{Binned $B$-band space densities for all galaxies (points) and the best-fitting maximumum likelihood Schechter function (smooth curves). Also shown are results from  \citet[][COMBO17]{bell04},  \citet[][DEEP2]{willm06},  \citet[][COSMOS]{zucca09} and  \citet[][AGES]{cool12}. Low redshift ($z \sim 0.1$) LFs from SDSS \citep{blant06} and 2dFGRS \citep[][]{madgw02} are also shown and the latter provides a fixed reference in each panel.  Significant fading of the bright end of the LF is evident.  As quantified later in Figures \ref{fig:B_phi_star} to \ref{fig:B_fig:Mfixed}, our LFs are in broad agreement with the literature, but smoother due to the large sample size.}	
	\label{fig:LF_binned_redandblue}
\end{figure}

\begin{figure}
 	\centering
	\includegraphics[width=0.45\textwidth]{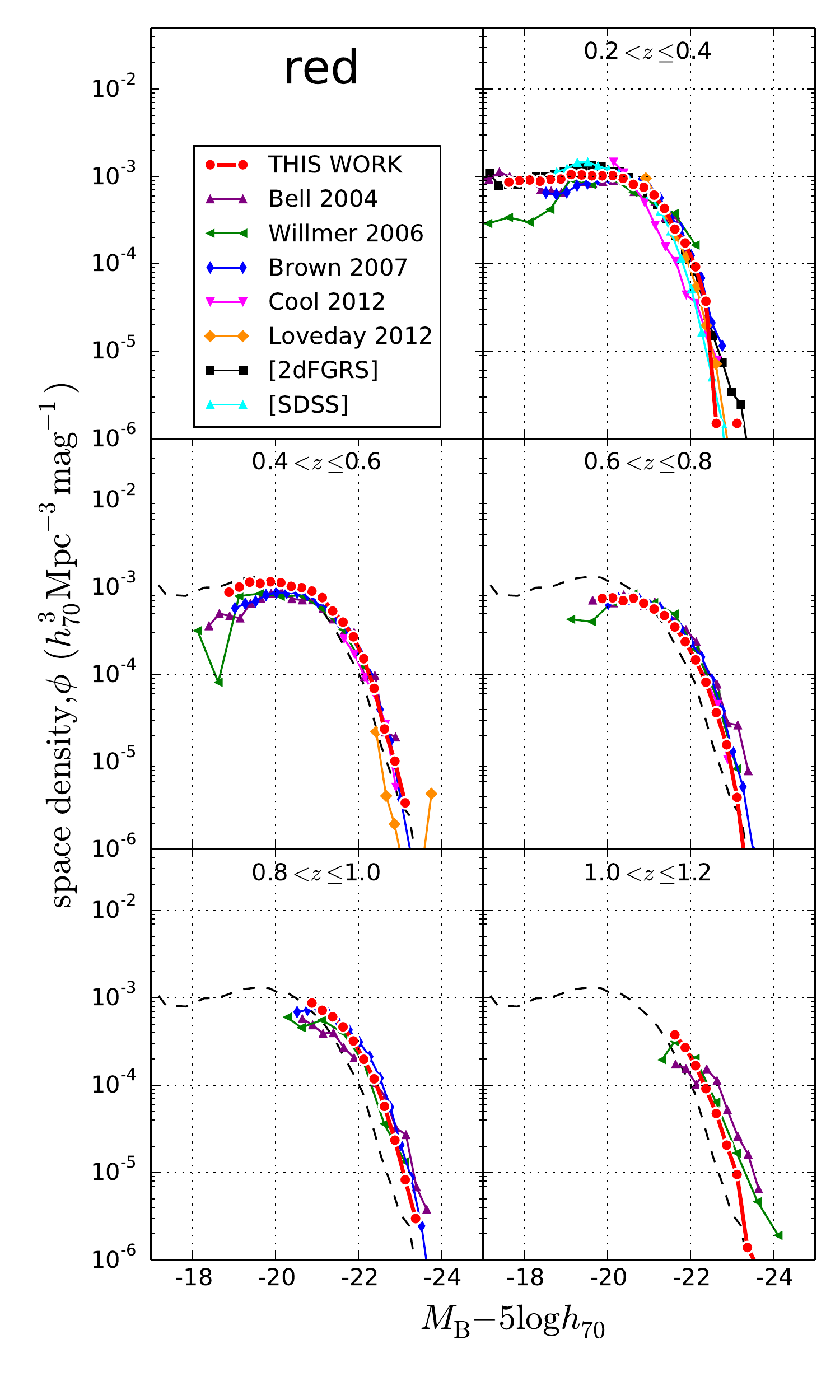}
	\caption{Binned $B$-band space densities for red galaxies (points) and the best-fitting maximumum likelihood Schechter function (smooth curves). Also shown are results from the literature, as detailed in the caption to Figure \ref{fig:LF_binned_redandblue}.  The low redshift LF of 2dFGRS galaxies \citep[][]{madgw02} provides a fixed reference in each panel. The fading of the bright end of the LF is clear, and largely accounted for by passive evolution as the text argues. The peak of the LF increases with decreasing redshift, indicating a build up of red galaxy stellar mass.  As quantified later in Figures \ref{fig:B_phi_star} to \ref{fig:B_fig:Mfixed}, our LFs are in broad agreement with the literature, but smoother due to the large sample size.}	
	\label{fig:LF_binned_red}
\end{figure}

\begin{figure}
 	\centering
	\includegraphics[width=0.45\textwidth]{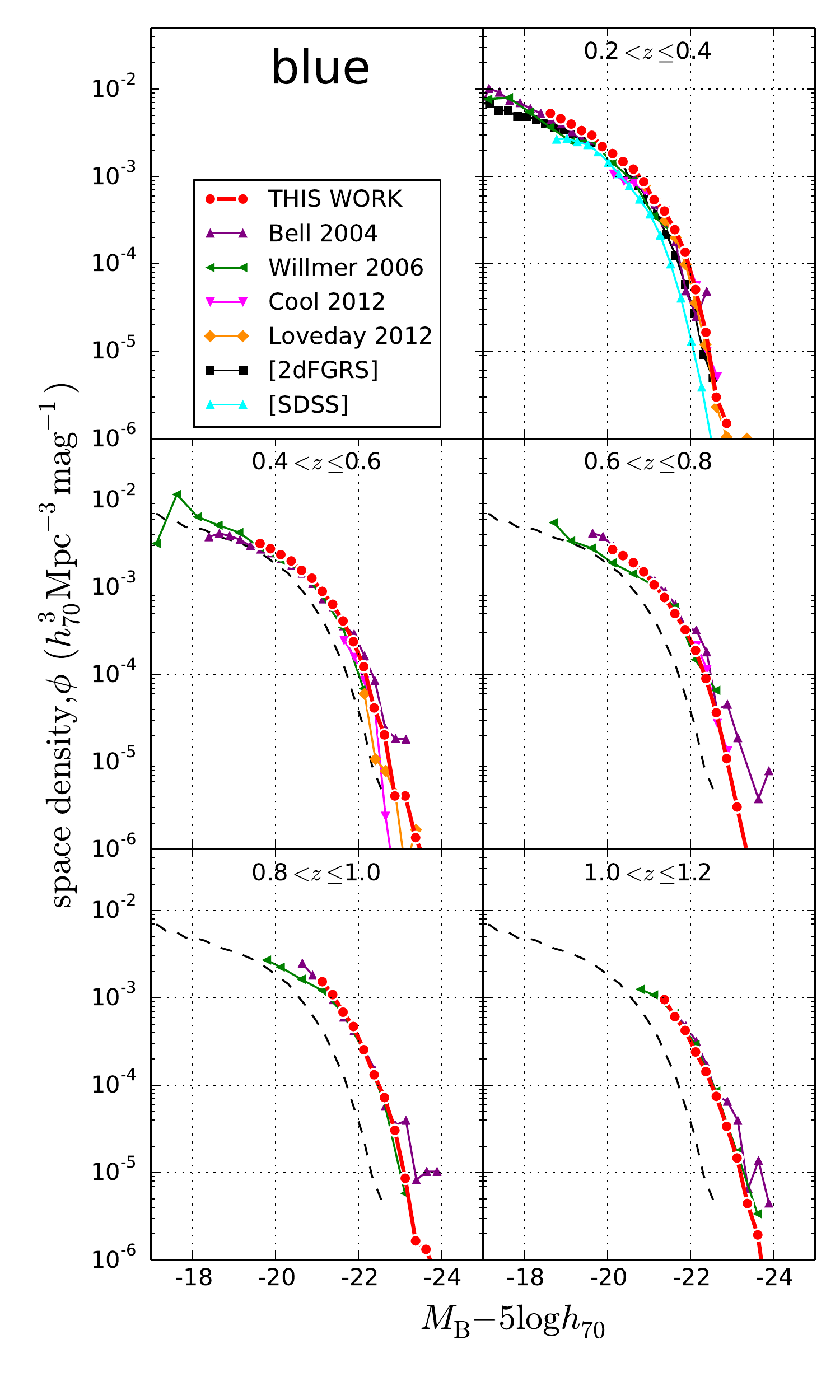}
	\caption{Binned $B$-band space densities for blue galaxies (points) and the best-fitting maximumum likelihood Schechter function (smooth curves). Also shown are results from the literature, as detailed in the caption to Figure \ref{fig:LF_binned_redandblue}.  The low redshift LF of 2dFGRS galaxies \citep[][]{madgw02} provides a fixed reference in each panel.  The fading of the bright end of the LF is clear, and largely accounted for by downsizing. As quantified later in Figures \ref{fig:B_phi_star} to \ref{fig:B_fig:Mfixed}, our LFs are in broad agreement with the literature, but smoother due to the large sample size.}	
	\label{fig:LF_binned_blue}
\end{figure}

\begin{figure}
 	\centering
	\includegraphics[width=0.45\textwidth]{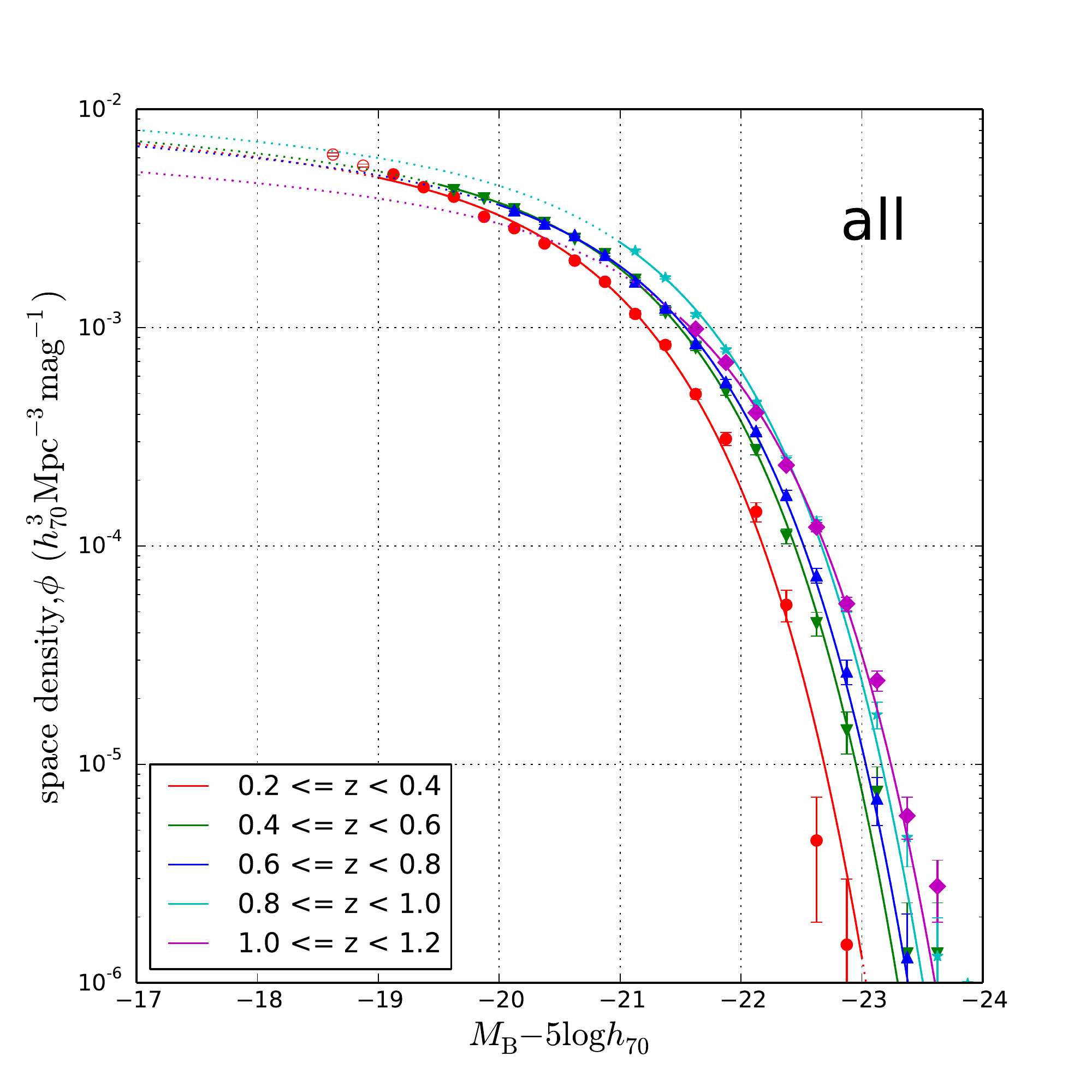}
	\caption{Evolution of the $B$-band luminosity function for all galaxies.  Best-fitting maximumum likelihood Schechter functions for different redshift bins are shown by smooth curves. Filled symbols indicate space densities included in the maximum likelihood fitting. Error bars show $1-\sigma$ Poisson errors for the binned space densities.}	
	\label{fig:evolution_redandblue}
\end{figure}

\begin{figure}
 	\centering
	\includegraphics[width=0.45\textwidth]{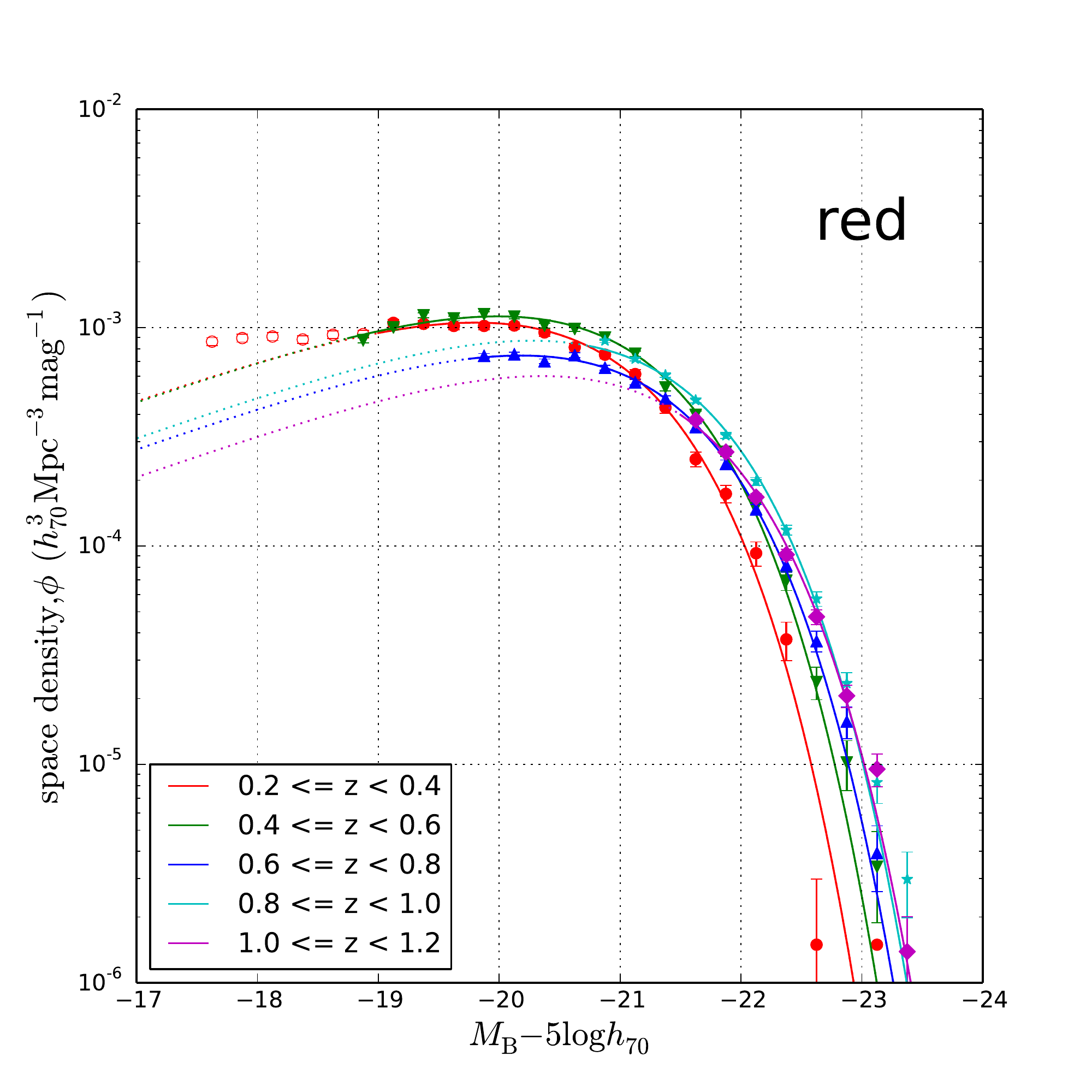}
	\caption{Evolution of the $B$-band luminosity function for red galaxies.  Best-fitting maximumum likelihood Schechter functions for different redshift bins are shown by smooth curves. Filled symbols indicate space densities included in the maximum likelihood fitting.  Open symbols show space densities for faint galaxies excluded from the fitting because of the excess density of red galaxies above a Schechter function at the faint end.  Error bars show $1-\sigma$ Poisson errors for the binned space densities.}	
	\label{fig:evolution_red}
\end{figure}

\begin{figure}
 	\centering
	\includegraphics[width=0.45\textwidth]{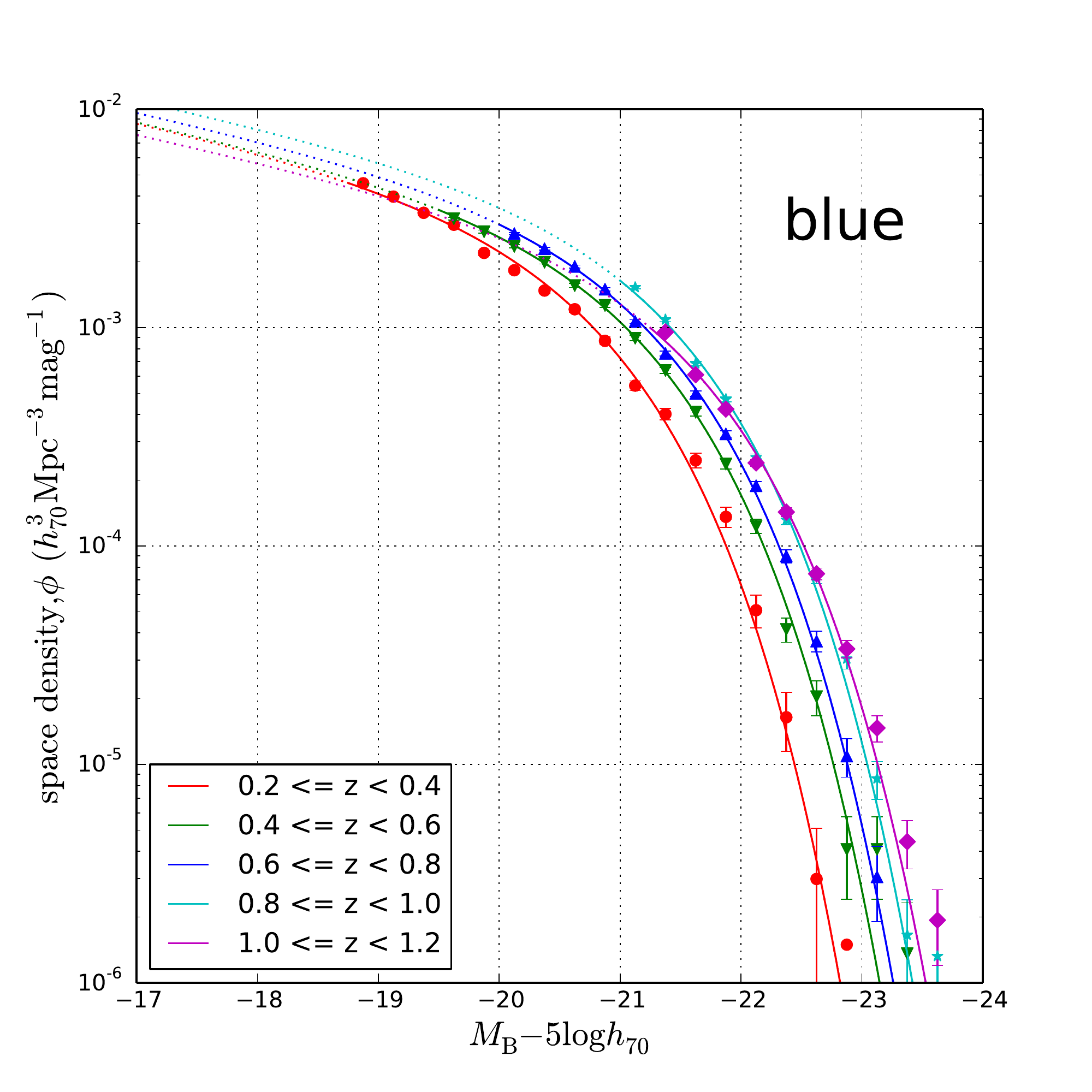}
	\caption{Evolution of the $B$-band luminosity function for blue galaxies.  Best-fitting maximumum likelihood Schechter functions for different redshift bins are shown by smooth curves. Filled symbols indicate space densities included in the maximum likelihood fitting.  Error bars show $1-\sigma$ Poisson errors for the binned space densities.}	
	\label{fig:evolution_blue}
\end{figure}

\clearpage

\section{Results and discussion} 
\label{sec:results}

\subsection{The evolution of space density and characteristic magnitude} 
\label{sec:results_spacedens}

Our binned space densities and maximum likelihood fits are shown in Figures \ref{fig:LF_binned_redandblue} to  \ref{fig:LF_binned_blue} and tabulated in Tables \ref{tab:bin_densities_redandblue} to \ref{tab:bin_densities_blue}. We plot the low redshift luminosity functions from 2dFGRS \citep{madgw02} in all bins to provide a fixed reference. Results from the prior literature are also shown in the figures for comparison purposes.

\begin{figure}
 	\centering
		\includegraphics[width=0.45\textwidth]{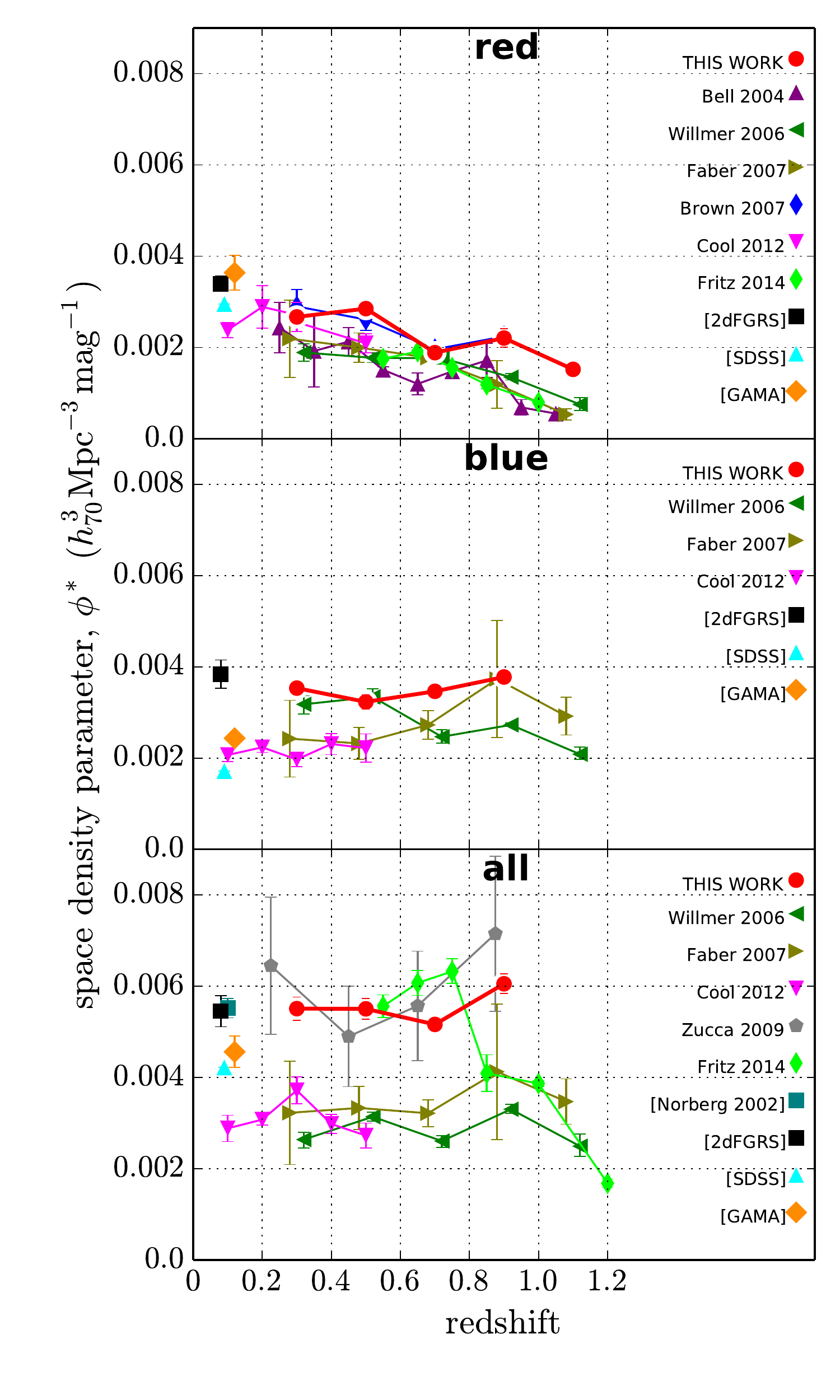}	
		\caption{Evolution of the $B$-band characteristic space density $\phi^*$, assuming fixed alpha values of $-0.5, -1.3$ and $-1.1$ for red, blue and all galaxies respectively. $\phi^*$ for red galaxies increases by \s50\% from $z=1.1$ to $z=0.3$ while $\phi^*$ for blue galaxies hardly changes from $z=0.9$ to $z=0.3$.  (We discount the points at $z = 1.1$ for blue and all galaxies because of photometric redshift uncertainty.)  Also shown are results from \citet[][COMBO17]{bell04}, \citet[][NDWFS]{brown07},  \citet[][COMBO17]{faber07},  \citet[][DEEP2]{willm06},  \citet[][COSMOS]{zucca09} and \citet[][AGES]{cool12}, and low redshift ($z\sim0.1$) results from  \citet[][2dFGRS]{madgw02}, \citet[][SDSS]{blant06} and \citet[][GAMA]{loved12}. As described in the text we calculated the SDSS Schechter parameters using Table 2 of \citet{blant06}, restricting ourselves to galaxies brighter than $M_B=-18.5$. Error bars on our results show errors due to cosmic variance. Error bars on results from the literature are as published (except those for SDSS which are not shown). As explained in Section \ref{sec:literature}, the low $\phi^*$ values for all galaxies from \citet{faber07, willm06} and \citet{cool12} can be ascribed to their adoption of a steeper faint slope parameter $\alpha$.}
		\label{fig:B_phi_star}
\end{figure}

\begin{figure}
 	\centering
		\includegraphics[width=0.45\textwidth]{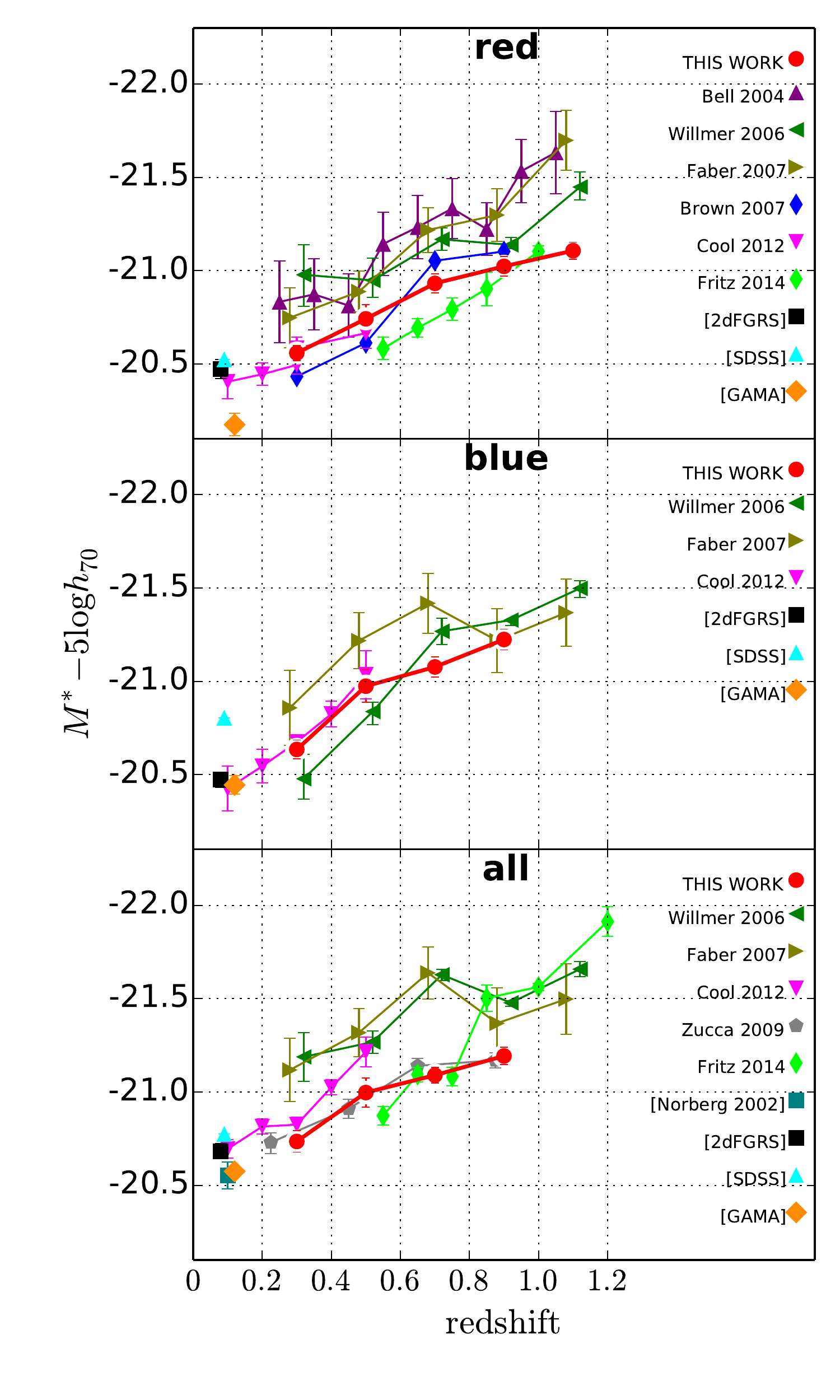}	
		\caption{Evolution of the $B$-band characteristic magnitude $M^*$, assuming fixed alpha values of $-0.5, -1.3$ and $-1.1$ for red, blue and all galaxies respectively. $M^*$ for red galaxies fades by 0.6 mag per unit redshift from $z=1.1$ to $z=0.3$ while that for blue galaxies fades more (0.8 mag per unit redshift) from $z=0.9$ to $z=0.3$.  (We discount the points at $z = 1.1$ for blue and all galaxies because of photometric redshift uncertainty.)  Also shown are results from the literature as listed in each panel and referenced in the caption to Figure \ref{fig:B_phi_star}.}
		\label{fig:B_M_star}
\end{figure}

Figures \ref{fig:evolution_redandblue} to  \ref{fig:evolution_blue} show the evolution of our maximum likelihood Schechter functions on single plots for all, red and blue galaxies.  Evolution of the corresponding Schechter parameters is given in Table \ref{tab:results_fixed} and shown graphically in Figures \ref{fig:B_phi_star} and \ref{fig:B_M_star}.

Blue galaxies are more numerous than red at all redshifts and the difference is particularly marked for faint galaxies, as red galaxies show a downturn in space density at faint magnitudes (per unit magnitude, but not per unit luminosity), whereas the space density of blue galaxies continues rising steeply towards fainter magnitudes. In the first redshift bin in Figures \ref{fig:LF_binned_red} and  \ref{fig:evolution_red} we can just detect the same upturn in the number of very faint red galaxies below $M_{\textrm{B}} \sim -19.0$ that has been reported by other authors \citep[e.g.][]{blant05, madgw02}.  This upturn represents an excess of very faint red galaxies above the numbers predicted by a pure Schechter function, and is generally parameterised by adding a second Schechter term \citep[e.g.][]{blant05, madgw02, loved12}. However, our measurements do not extend to faint enough magnitudes to measure the upturn in faint galaxy numbers in the same way that can be done in the low redshift Universe.

We see from Figures \ref{fig:LF_binned_redandblue} to \ref{fig:evolution_blue} and Figure \ref{fig:B_M_star} that that the overall distribution of luminosities is fading, with the blue galaxy distribution fading faster than the red. The fading of 0.6 mag per unit redshift in the values of $M^*$ for red galaxies from $z = 1.1$ to $z = 0.3$ is due to a combination of passive stellar fading and the arrival of new galaxies on the red sequence as they cease to form stars. The fading of 0.8 mag per unit redshift for blue galaxies from $z = 0.9$ to $z = 0.3$ is consistent with downsizing \citep{cowie96} -  that on average more massive galaxies in the blue cloud cease star formation and move across the green valley to the red sequence earlier than less massive ones. (We omit the highest redshift bin for blue and all galaxies because redshift uncertainty makes fitting a Schechter function unreliable when the faint end slope is steep and space density measurements are not available for fainter galaxies.)

The characteristic space density $\phi^*$ (shown in Figure \ref{fig:B_phi_star}) provides an approximate measure of the space density close to the characteristic magnitude ($\phi = 1.086\phi^*$ at $M = M^ *$). For red galaxies $\phi^*$ increased by \s50\% from $z \sim 1.1$ to $z \sim 0.3$ while $\phi^*$ for blue galaxies changed very little from $z \sim 0.9$ to $z \sim 0.3$. These trends are also consistent with the migration of blue galaxies to the red sequence and with downsizing.

A detailed interpretation of $M^*$ and $\phi^*$ evolution is not straightforward. This is partly due to the well-known degeneracy between the Schechter parameters, but it also due to the complexity of the various physical processes involved in transforming the properties and space density of galaxies, especially blue galaxies.

\vspace{20 pt}

\subsection{The evolution of luminosity density} 
\label{sec:results_lumdens}

	Luminosity density $j$ has a more direct physical interpretation than the individual Schechter parameters $M^*$ and $\phi^*$ as it represents the total flux emitted by all the stars in a galaxy population in a particular waveband. We use Equation \ref{eq:lumdens} to determine luminosity density from the Schechter parameters for each redshift bin, integrating over all luminosities from zero to infinity. For blue galaxies, $\alpha \lesssim -1.0$ and the Schechter function increases without limit as $L\to 0$, but the faintest galaxies do not contribute significantly to the total luminosity density. However the total luminosity density as given by Equation \ref{eq:lumdens} does depend sensitively on the precise value of $\alpha$. For example, with a typical characteristic magnitude $M^* = -20.5$ we find that the fraction of the luminosity contributed by galaxies fainter than $M^* = -17$ is $12\% (6\%, 1\%)$ for $\alpha = -1.3$ $(-1.1, -0.5)$.
	
For red galaxies, $\alpha \sim -0.5$ and the space density decreases at fainter magnitudes so that it is insensitive to the precise value of $\alpha$ adopted, and this is reflected in the fact that $\Gamma(\alpha + 2)$ in Equation \eqref{eq:lumdens} has a minimum at $\alpha = -0.5$. As already indicated, several studies have detected an excess of very faint red galaxies above the predictions of a simple Schechter function  model, but despite this excess, the number of red galaxies still decreases so rapidly as $L$ decreases to the faintest luminosities that we do not introduce significant error in the computed luminosity density by excluding it from our calculations.
	
\begin{figure}
 	\centering
		\includegraphics[width=0.45\textwidth]{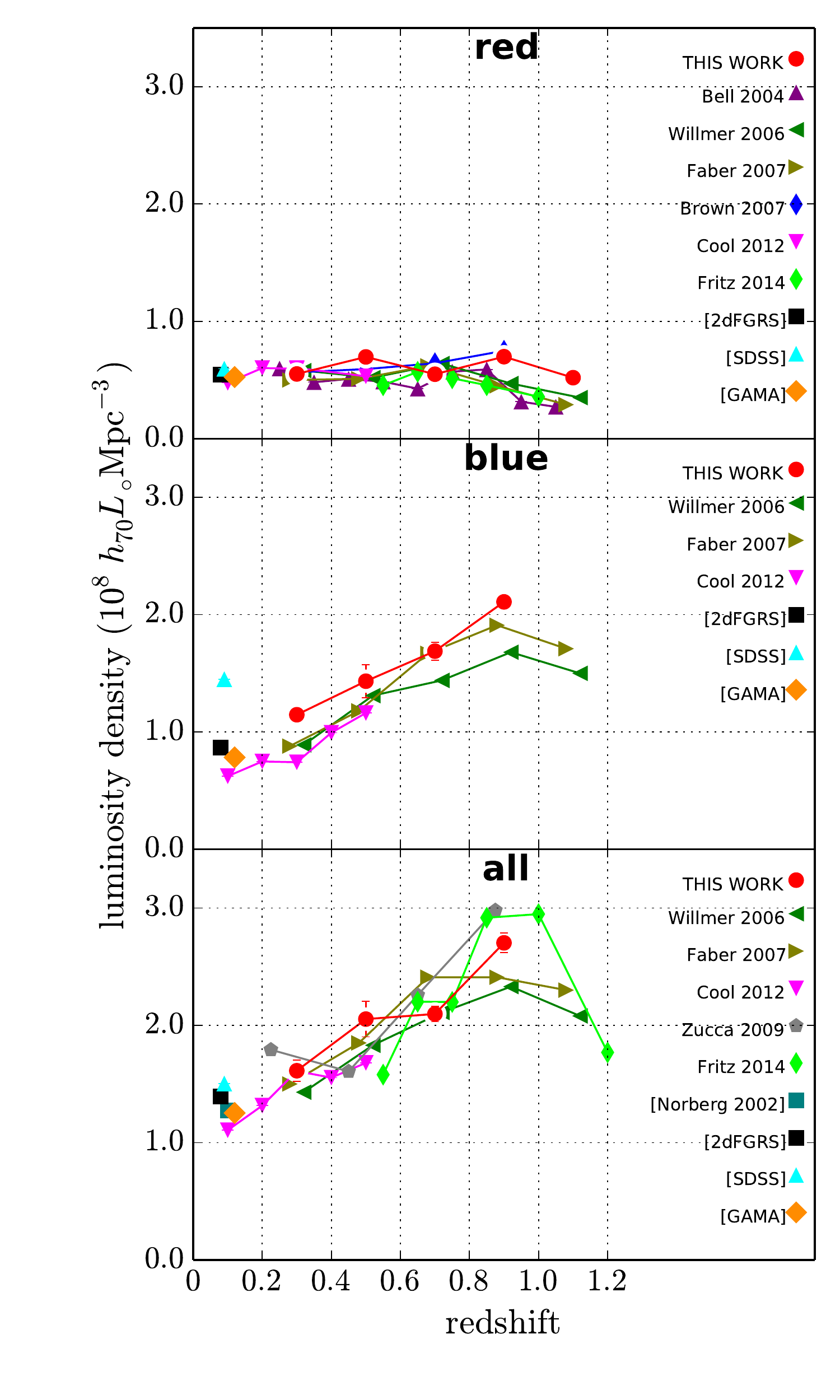}	
		\caption{Evolution of the $B$-band luminosity density $j$, assuming fixed alpha values of $-0.5, -1.3$ and $-1.1$ for red, blue and all galaxies respectively. The luminosity density of red galaxies increases marginally from $z=1.1$ to $z=0.3$ while that for blue galaxies almost halves from $z=0.9$ to $z=0.3$.  (We discount the points at $z = 1.1$ for blue and all galaxies because of photometric redshift uncertainty.)  Also shown are results from the literature as listed in each panel and referenced in the caption to Figure \ref{fig:B_phi_star}. In the case of SDSS and where stellar luminosity densities have not been quoted by the author we have determined them from the Schechter parameters using Equation \eqref{eq:lumdens}.}	
		\label{fig:B_lumdens}
\end{figure}

As Figure \ref{fig:B_lumdens} shows, we find that the total $B$-band stellar luminosity density of red galaxies increased marginally from $z \sim1.1$ to $z \sim0.3$ while that of blue galaxies almost halved from $z \sim0.9$ to $z \sim0.3$. (Again we omit the highest redshift bins for blue and all galaxies because redshift uncertainty makes determination of  Schechter function parameters unreliable.)

For red galaxies, luminosity density provides a relatively good proxy for stellar mass, as stellar mass to light ($M/L$) ratios correlate well with optical colors  \citep[e.g.][]{bell01, bell03, taylo11, wilki13} and these vary little from one red galaxy to another. Furthermore, the red galaxy luminosity density is insensitive to the adopted value of $\alpha$. Most of the red galaxy luminosity density  comes from galaxies close to $M^*$ (e.g. $\sim 80\%$ from galaxies within 1.2 mag of $M^*$ for $\alpha = -0.5$, $M^* = - 20.0$).  

\begin{figure}
 	\centering
		\includegraphics[width=0.45\textwidth]{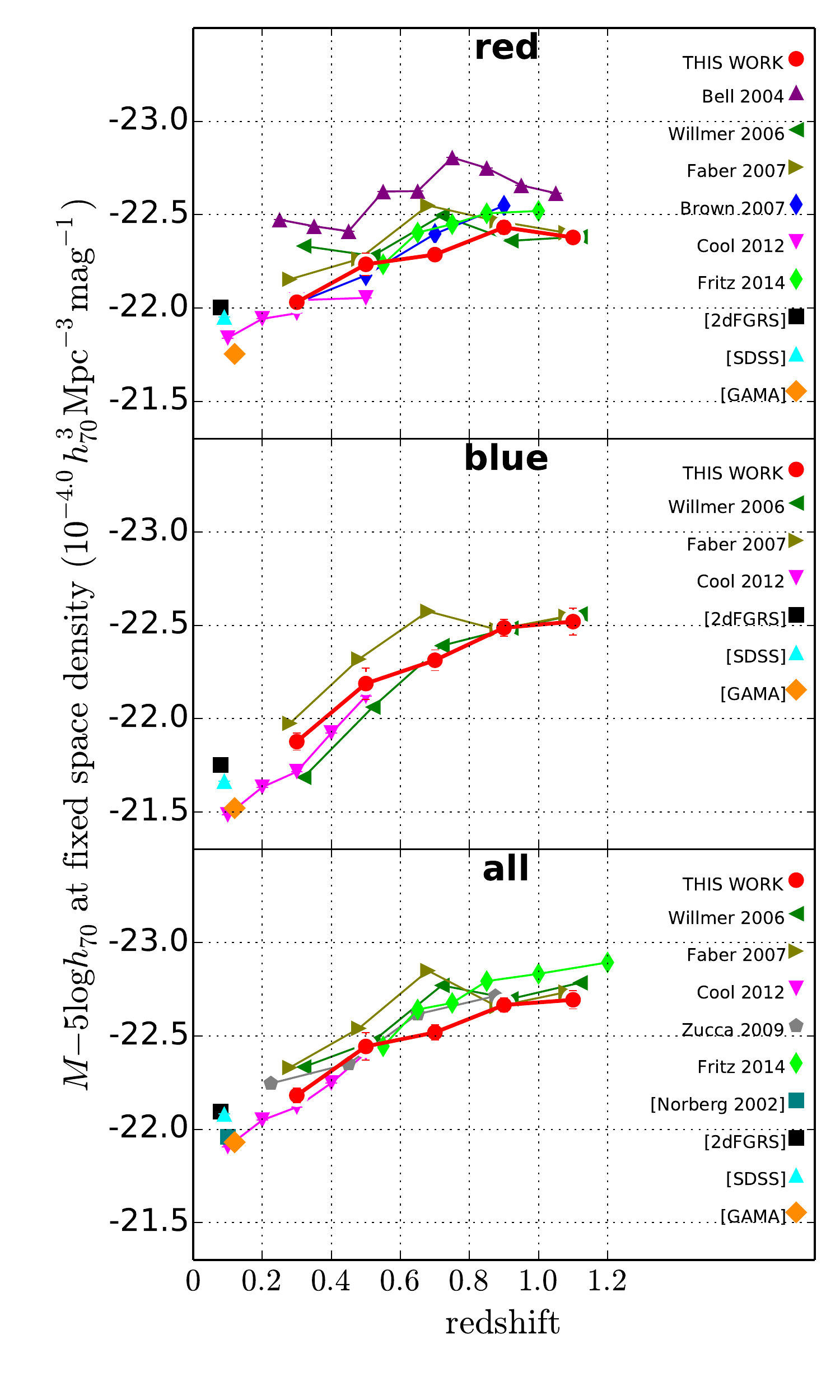}	
		\caption{Evolution of the bright end of the $B$-band luminosity function from $z=1.1$ to $z=0.3$, assuming fixed alpha values of $-0.5, -1.3$ and $-1.1$ for red, blue and all galaxies respectively. The luminosity evolution of the brightest galaxies is indicated by the value of $M_{\rm{B}} - 5 \log h_{\rm{70}}$ at which the space density is $10^{-4.0} h_{\rm{70}}^3 \, \rm{ Mpc}^{-3} \, \rm{mag}^{-1} $.  The rate of fading of individual highly luminous red galaxies increased from $z=1.1$ to $z=0.1$.  Also shown are results that we have calculated from the Schechter parameters published in the optical LF literature  as listed in each panel and referenced in the caption to Figure \ref{fig:B_phi_star}.}
		\label{fig:B_fig:Mfixed}
\end{figure}

\begin{figure}
 	\centering
		\includegraphics[width=0.4\textwidth]{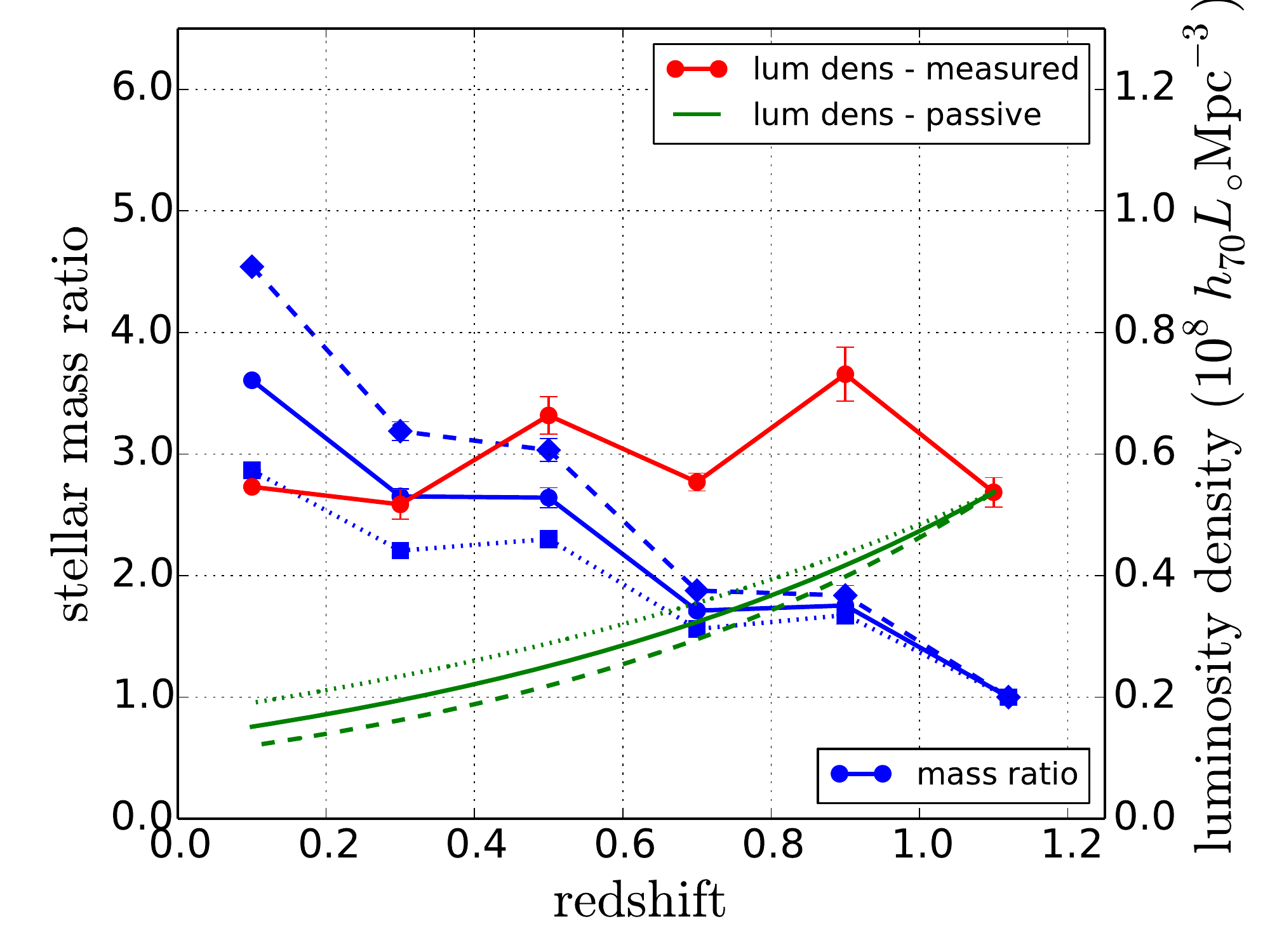}	
		\caption{The luminosity density of red galaxies has changed relatively little from $z = 1.1$ to $z = 0.1$ (\textit{right axis and red points}).  The point at $z = 0.1$ is for 2dFGRS \citep{madgw02}. Comparison with evolution of a passive stellar population whose stellar mass to light ratio has evolved according to $d\log(M/L)/dz = -0.55$ (\textit{right axis and solid green curve})  implies that the stellar mass in red galaxies increased by a factor of $\sim 3.6$ (\textit{left axis and blue points}). The dashed and dotted curves show the results assuming faster and slower rates of passive evolution with $d\log(M/L)/dz = -0.65$ and -0.45 respectively. Error bars on our results show errors due to cosmic variance.}
		\label {fig:B_luminosity_density_red}
\end{figure}

\begin{figure}
 	\centering
		\includegraphics[width=0.45\textwidth]{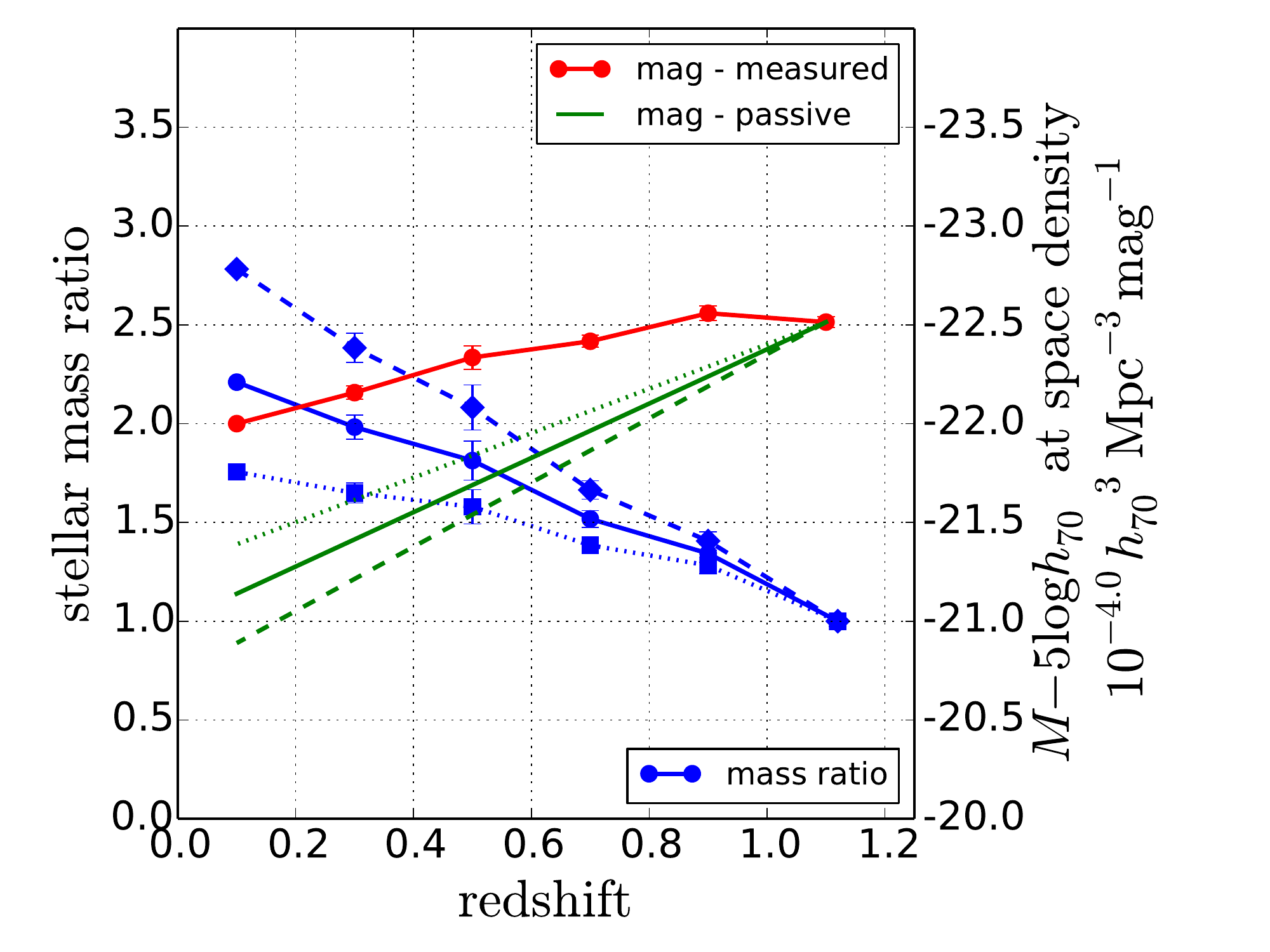}	
		\caption{The absolute magnitude of luminous red galaxies at a fixed comoving space density of $10^{-4.0} h_{70}^3 \, \rm{ Mpc}^{-3} \, \rm{mag}^{-1} $ \textit{(right axis and red points)} is seen to fade at an increasingly rapid rate from $z = 1.1$ to $z = 0.1$.  The point at $z = 0.1$ is for 2dFGRS \citep{madgw02}. By comparing this rate of fading with that to be expected on the basis of pure luminosity evolution with $d\log(M/L)/dz = -0.55$ (\textit{right axis and solid green line}), we find that individual highly luminous red galaxies slightly more than doubled in mass from $z = 1.1$ to $z \sim 0.1$ (\textit{left axis and blue points}).  The dashed and dotted curves show the results assuming faster and slower rates of passive evolution with $d\log(M/L)/dz = -0.65$ and -0.45 respectively. Error bars on our results show errors due to cosmic variance.}
		\label {fig:highly_luminous_red}
\end{figure}

\begin{figure}
 	\centering
		\includegraphics[width=0.49\textwidth]{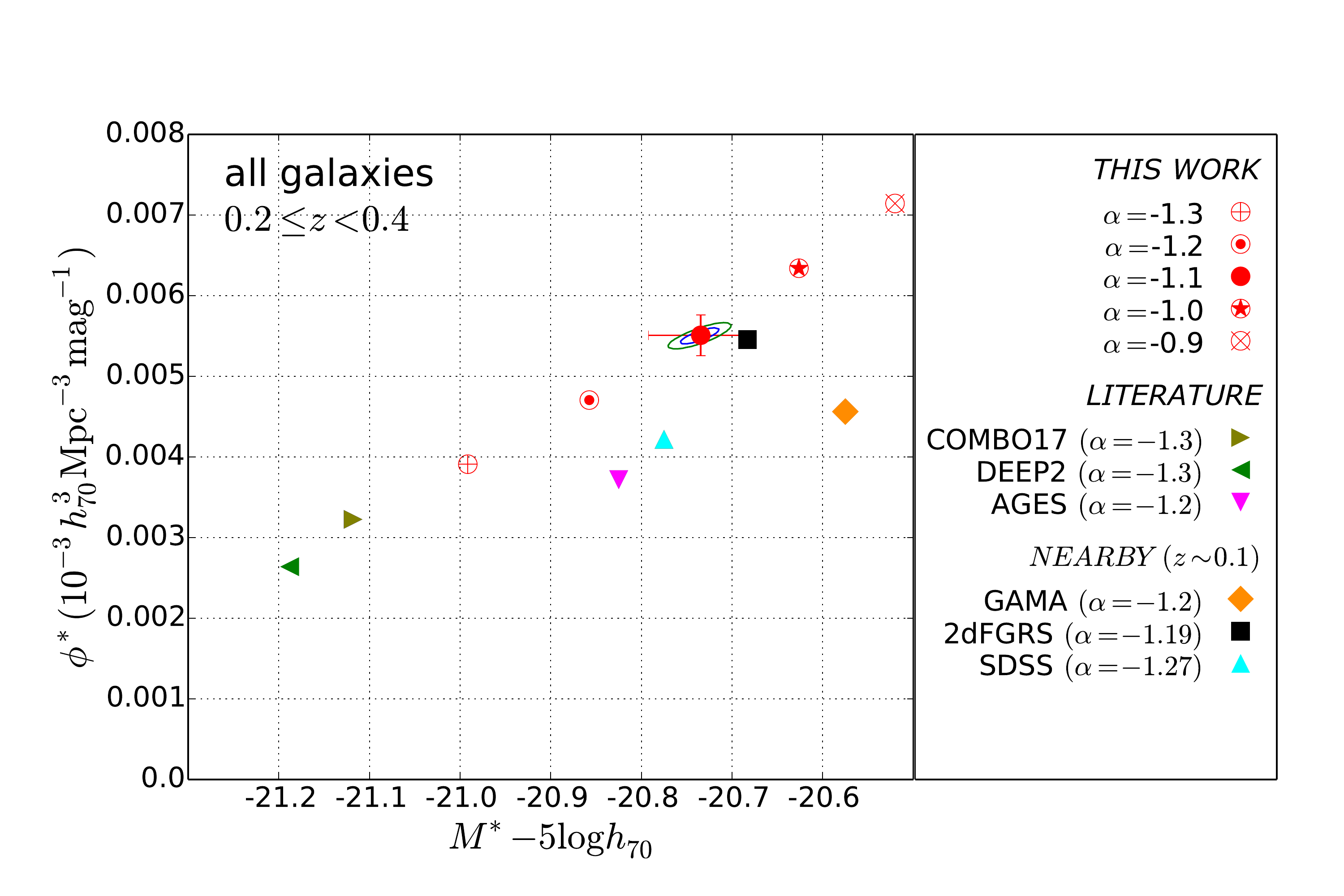}	
		\caption{Showing that maximum likelihood values of $\phi^*$ and $M^*$ are critically dependent on the fixed value of $\alpha$ adopted. $(M^*, \phi^*)$ values are plotted as red circles for all galaxies at $0.2 \leq z < 0.4$ and five different values of  $\alpha$: our preferred value of $\alpha = -1.1$, and four others.  The error bars for our results with $\alpha = -1.1$ indicate likely random error due to cosmic variance estimated using subfields, while the contours show 68\% and 95\% confidence limits for the maximum likelihood fit.}
		\label {fig:phistar_Mstar_errors}
\end{figure}

\begin{figure}
 	\centering
		\includegraphics[width=0.49\textwidth]{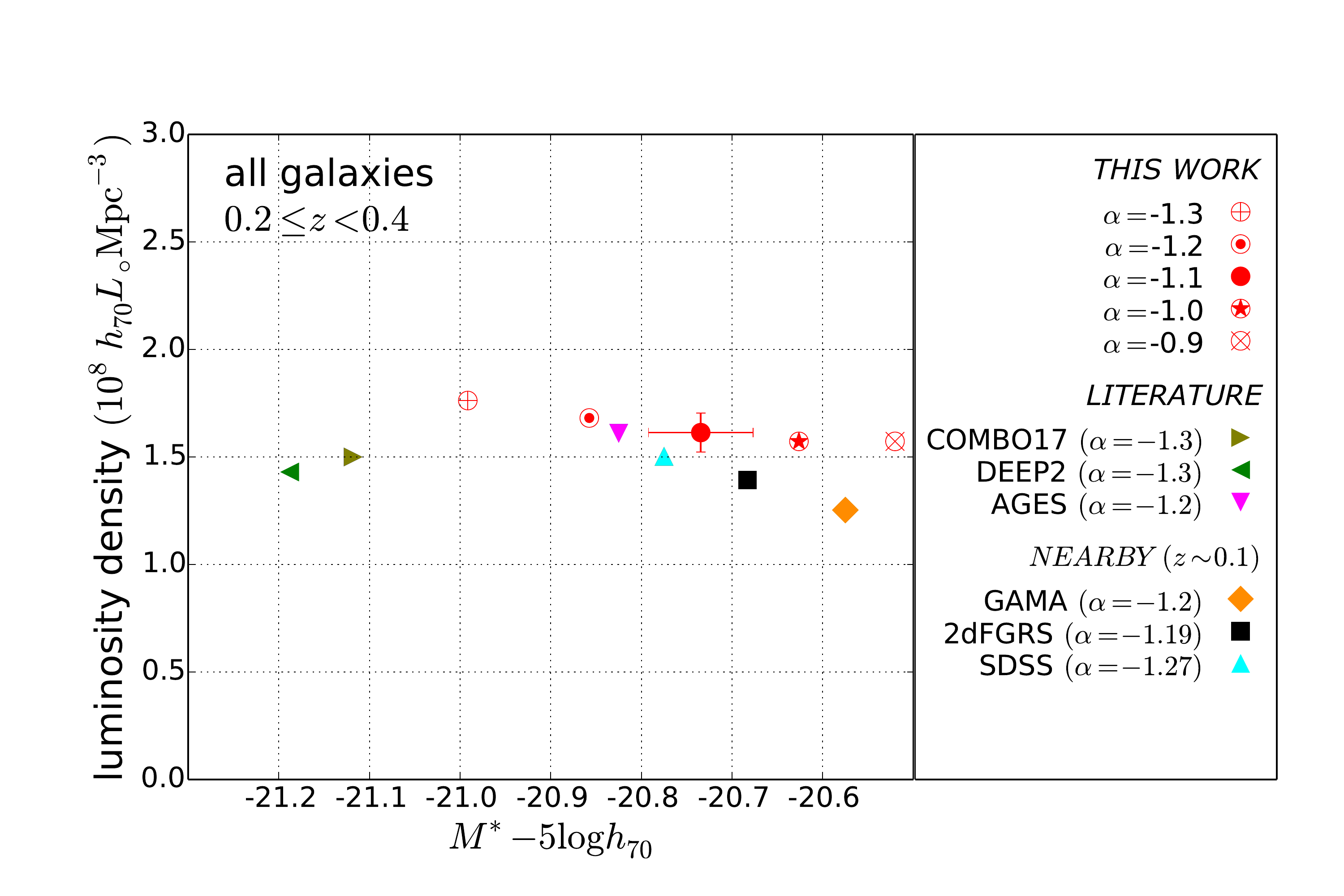}	
		\caption{Showing that luminosity density measurements are affected relatively little by the fixed value of $\alpha$ adopted, even though the maximum likelihood values of $\phi^*$ and $M^*$ vary considerably, as shown in Figure \ref{fig:phistar_Mstar_errors}. Similar plots show that the measured luminosity density of red galaxies is much less dependent on $\alpha$ than that of blue.} 		
		\label {fig:lumdens_Mstar_errors}
\end{figure}

Red galaxy stellar luminosity density has not faded as fast as it would have done due to passive stellar evolution alone, and we attribute the difference almost entirely to galaxies migrating from the blue cloud, with no contribution from mergers because dry mergers between quiescent red galaxies do not change the total red galaxy luminosity density. 

 We can estimate how the stellar mass in red galaxies evolves if we assume a fixed value $G \, (<0)$ for the rate of change with redshift of $\log( M/L)$ for quiescent galaxies:

\begin{equation}\label{eq:MtoL}
	\frac{d}{dz} \log \left( \frac{M}{L} \right) = G.
\end{equation}

In terms of total stellar mass density $m$ and total luminosity density $j$ for the stellar populations in red galaxies this becomes:

\begin{equation}\label{eq:MtoL_lumdens}
	\frac{d}{dz} \log \left( \frac{m}{j} \right) = G.
\end{equation}

 Given measurements of the initial luminosity density $j_0$ at redshift $z_0$ and the luminosity density $j$ at any subsequent redshift $z$, we can subtract the contribution to luminosity density evolution due to passive evolution, and estimate how much the stellar mass density has increased. From Equation \eqref{eq:MtoL_lumdens}:

\begin{equation}\label{eq:mass_ratio}
	\log \left(\frac{m}{m_0}\right) = \log \left(\frac{j}{j_0}\right) + G(z - z_0) .
\end{equation}

We adopt the value $G = -0.55$ for $d\log(M/L)/dz$ based on evolution of single burst SSPs and on studies of the evolution of the Fundamental Plane up to $z \sim 1$. Values derived from evolution of single burst \citet{bruzu03} SSP models (solar metallicity, Chabrier IMF, Padova 1994 library) are -0.52, -0.55, -0.61 and -0.71 for star formation redshifts $z_f$ of 4, 3, 2, 1 respectively. Our value of $G = -0.55$ corresponds to $z_f = 3$ and differs by only 5\% from the $z_f = 4$ value and 10\% from the $z_f = 2$ value.

Values for $G$ derived from evolution of the Fundamental Plane are -0.72 for field spheroidals in \citet{treu05}, -0.66 for early type galaxies in \citet{wel06}, -0.60 for cluster galaxies in \citet{holde10}, -0.54 for cluster galaxies in \citet{sagli10} and -0.76 for field galaxies in \citet{sagli10}. Given the large scatter in Fundamental Plane $M/L$ measurements, especially for less massive galaxies \citep[e.g.][]{treu05, sagli10} these slightly faster rates of evolution agree with those derived from SSP models within the measurement errors.
 
 The assumed rate of fading represents a luminosity weighted average value over red galaxies of all masses and our analysis does not take into account the fact that star formation peaked earlier in more massive galaxies \citep[e.g.][]{deluc06, thoma10, mores10} resulting in slower passive evolution in highly luminous massive galaxies \citep[e.g.][]{treu05, sagli10}.  Our simplified approach enables us to obtain an approximate measure of stellar mass growth in red galaxies. An alternative approach uses the tight correlation of optical $M/L$ ratios with restframe optical colors \citep[e.g.][]{bell01, bell03, taylo11, wilki13} to measure evolution of the stellar mass function and many studies have done this \citep[][]{drory05, bundy06, borch06, arnou07, perez08, ilber10, bramm11, gonza11, mortl11, ilber13, moust13, muzzi13b}. We take this approach in Paper II. Although they have the advantage of being conceptually simple, measurements of SMF evolution suffer from the disadvantage that they are dependent on the particular choices of model used (e.g. SED fit or $M/L$ ratio, dust obscuration, stellar IMF). By contrast LF evolution measurements are model independent, and the simple conclusions presented here regarding red galaxy stellar mass evolution can easily be modified to take account of any more precise future measurements of stellar $M/L$ ratios.

Taking the stellar mass as unity in arbitrary units at $z_0 = 1.1$, Equation \eqref{eq:mass_ratio} enables us to determine how the stellar mass density in red galaxies has evolved. As Figure \ref{fig:B_luminosity_density_red} shows, we find that overall the stellar mass in red galaxies increased by a factor of around 3.6 from $z \sim 1.1$ to $z \sim 0.1$.  Increasing the rate of passive fading $-d\log(M/L)/dz$ by 0.1 (i.e. 15\% or 0.25 mag per unit redshift) increases this factor to 4.5, while decreasing it by a similar amount reduces it to 2.9.

\subsection{The evolution of highly luminous galaxies} 
\label{sec:results_luminous}

Because of the steepness of the bright end of the luminosity function, a small amount of evolution in galaxy luminosity and small photometric errors  can produce large changes in the space density at fixed luminosity. However, it is possible to accurately measure the evolution of the magnitude $M^{\rm{fixed}}$ corresponding to a fixed space density, and we choose to do this for a space density of $10^{-4.0} h_{70}^3 \, \rm{ Mpc}^{-3} \, \rm{mag}^{-1} $. Our results are given in Table \ref{tab:results_fixed} and plotted in Figure \ref{fig:B_fig:Mfixed}, together with results based on Schechter parameters from the literature. We find that the most massive red galaxies are $\sim0.4$ mag fainter at $z \sim 0.3$ than at $z \sim 1.1$, while highly luminous blue galaxies are $\sim0.5$ mag fainter at $z \sim 0.3$ than at $z \sim 0.9$. The rate of fading has been increasing in both cases.

As discussed in Section \ref{sec:LFintro}, \citet{bell04} demonstrated that there were insufficient highly luminous blue galaxies at $z\sim 1$ to give rise to the highly luminous red galaxies we see at lower redshifts via cessation of star formation. There have also been too few major mergers between red galaxies since $z\sim 1.0$ to account for their formation. The observed evolution in $M^{\rm{fixed}}$ for red galaxies must therefore be due to a combination of passive evolution and minor mergers in a fixed population of massive red galaxies. As can be seen from Figure \ref{fig:B_fig:Mfixed}, massive, highly luminous red galaxies have been fading at an increasing rate since $z \sim 1$. Including the results from 2dFGRS at $z\sim0.1$, we find that for individual highly luminous red galaxies the rate of fading increased from \s0.2 mag per unit redshift at $z = 1.0$ to \s0.8 at $z = 0.2$.

As with total stellar mass density in the previous section, we can make allowance for the passive fading of the stars in highly luminous red galaxies and estimate their increase in mass due to minor mergers. As discussed in the previous section, today's massive red galaxies formed their stars earlier than less massive ones and have therefore faded more slowly since $z\sim1$ than red galaxies as a whole. We do not take this into account but adopt the same preferred value for $G = d\log(M/L)/dz$ of -0.55 per unit redshift and indicate how varying this figure by $\pm0.1$ alters our conclusions. Writing $M$ for the mass of an individual luminous red galaxy and $M_B$ for its absolute $B$-band magnitude, Equation \eqref{eq:MtoL} becomes:

\begin{equation}\label{eq_highly_luminous}
	\frac{d}{dz} \left( \log M + 0.4M_B \right) = G.
\end{equation}

Taking the stellar mass of a luminous red galaxy as unity in arbitrary units at $z_0 = 1.1$, Equation \eqref{eq_highly_luminous} enables us to estimate the rate of mass increase due to mergers. As Figure \ref{fig:highly_luminous_red} shows, we find that the stellar mass in individual highly luminous red galaxies increased by a factor of around 2.2 from $z \sim 1.1$ to $z \sim 0.1$. Increasing the rate of passive fading $-d\log(M/L)/dz$ by 0.1 (i.e. 15\% or 0.25 mag per unit redshift) increases this factor to 2.8, while decreasing it by a similar amount reduces it to 1.8.

The situation for highly luminous blue galaxies cannot easily be interpreted, as new star formation and accretion by mergers can both produce brightening, while passive fading and the reduction or cessation of star formation can both result in fading. 

\subsection{Comparison with the literature } 
\label{sec:literature}

As can be seen from Figures \ref{fig:LF_binned_redandblue} to \ref{fig:LF_binned_blue} and \ref{fig:B_phi_star} to \ref{fig:B_fig:Mfixed}, our results are for the most part in broad agreement with previous authors.  In particular, our results line up well with those from low redshift surveys including 2dFGRS, SDSS and GAMA. It is particularly noticeable in Figures \ref{fig:B_phi_star} to \ref{fig:B_fig:Mfixed} that we see less scatter with redshift than other studies and we attribute this to our much larger sample size.

It is well known that $\phi^*$, $M^*$ and $\alpha$ are highly degenerate parameters.  In particular, the measured values of $\phi^*$ and $M^*$ depend critically on the value of $\alpha$ adopted. As Figure \ref{fig:B_phi_star} shows, our values for $\phi^*$ for red and blue galaxies combined are almost double those obtained by \citet{bell04, willm06} and \citet{faber07} using DEEP2 and COMBO17 data. However,  the discrepancy largely disappears if we adopt their value $\alpha = -1.3$ rather than our preferred value of  $\alpha = -1.1$, as can be seen from Figure \ref{fig:phistar_Mstar_errors}, which compares the $\phi^*$ and $M^*$ values obtained using $\alpha$ values of $-1.3, -1.2, -1.1, -1.0$ and $-0.9$ for the example redshift bin $0.2 \leq z < 0.4$. Figure \ref{fig:lumdens_Mstar_errors} shows that luminosity density measurements are affected relatively little by the fixed value of $\alpha$ adopted, even though the maximum likelihood values of $\phi^*$ and $M^*$ vary considerably. This is fortunate as luminosity density is a more physically meaningful quantity than the individual Schechter parameters.  

We calculated the SDSS Schechter parameters using the binned data from Table 2 of \citet{blant06}, restricting ourselves to galaxies brighter than $M_B=-18.5$. We did this because we could not obtain a satisfactory fit at the bright end if we included the fainter galaxies in the table (luminosities $M_B>-18.5$), and because the space density for all galaxies fainter than than $M_B\sim-18.5$ showed showed a significant downturn as compared with the measurements of \citet{blant05} for very near ($z<0.01$) SDSS galaxies which extend to much fainter magnitudes.

Cosmic variance uncertainties determined using subsamples are shown by error bars in Figure \ref{fig:phistar_Mstar_errors}. Uncertainties in performing the maximum likelihood fits are shown by $1-\sigma$ and $2-\sigma$ contours and these are comparable in magnitude. As one progresses to higher redshifts the maximum likelihood uncertainty decreases (because the numbers of galaxies in a redshift bin increases), while the cosmic variance error remains important. 

We find that the stellar mass in red galaxies as a whole increased  by a factor of around 3.6 from $z \sim 1.1$ to $z \sim 0.1$. Previous studies based on optical LFs have reported that it has at least doubled from $z \sim 1$ to $z \sim 0$ \citep[e.g.][]{bell04, brown07, faber07}. Studies based on SMFs have produced similar results. \citet{muzzi13b} measured an increase of 2.0 times in the stellar mass density of $M>10^{8}M_{\sun}$ quiescent galaxies from $1.0<z<1.5$ to $0.5<z<1.0$ and 1.7 times from $0.5<z<1.0$ to $0.2<z<0.5$.

\citet{brown07} used a similar method to this work in order to estimate the growth of stellar mass in red galaxies, reporting a stellar mass density increase of approximately 2 since $z = 1$. They assumed a slightly lower rate of passive fading based on a \citet{bruzu03} model of an SSP with $z_{\rm{form}} = 4$, $\tau = 0.6$ Gyr. This gave them fading of 1.24 mag per unit redshift which is equivalent to $G=d\log(M/L)/dz = -0.5$. Using our data this gives a mass density increase of 3.2 times.  \citet{brown07} also found that 80\% of the stellar mass in highly luminous red galaxies was already in place at $z = 0.7$ so that the stellar mass growth since then has been only 25\%.
 
\citet{cimat06} reanalysed COMBO-17 data \citep{bell04} and DEEP2 data \citep{faber07} also carrying out a comparison with pure luminosity evolution. They found that evolution of early type galaxies is strongly dependent on their stellar mass, with $M>10^{11}M_{\sun}$ galaxies having changed little in number density since $z\sim0.8$ but less massive galaxies having become more numerous. They suggested that at any redshift there is a critical mass above which virtually all stellar mass is already in place.

All these results are consistent with a model in which star-forming galaxies in the blue cloud cease to form stars and move across the green valley onto the red sequence causing a build up of quiescent SMD. They are also consistent with downsizing \citep{cowie96}:  that on average more massive galaxies in the blue cloud cease star formation and move across the green valley to the red sequence earlier than less massive ones. The recent study by \citet{ilber13} showed that the low mass end of the SMF of star-forming galaxies has evolved more rapidly from $z=4$ to $z = 0.2$ than the high mass end, indicating that SF in $M >\sim 10^{10.8} \, M_{\sun}$ galaxies has been quenched more rapidly than in less massive galaxies.  \citet{bundy06} showed that downsizing depends little on environment, except for the most massive galaxies, implying that it is governed by internal rather than external processes. 

Measuring the mass growth of massive galaxies has proved problematic historically, and this is reflected in the scatter between different authors seen in Figure \ref{fig:B_fig:Mfixed}. Our large sample size and use of the magnitude at fixed space density have allowed us to obtain more reliable measurements than hitherto. For individual highly luminous red galaxies we find that stellar mass has increased by a factor of \s2.2 from $z \sim 1.1$ to $z \sim 0.1$ and we see that the rate of mass assembly has been decreasing since $z \sim 1.1$, with \s90\% being assembled prior to $z = 0.5$. Previous authors have found that $~80\%$ of the stellar mass in $z \sim 0$ massive red galaxies was already in place by $z \sim 1$ \citep[e.g.][]{mortl11, brown07}. \citet{pozze10} found that the majority of massive ($M>10^{11}M_{\sun}$) early type galaxies were already in place at $z=1$, while  \citet{ilber13} found that the high mass end of the SMF evolved no more than 0.2 dex at $z<1$, implying that 60\% of the stellar mass was already in place at  $z \sim 1$. \citet{lin13} found that brightest cluster galaxies increased in mass by a factor of 2.3 from  $z = 1.5$ to  $z = 0.5$ with little growth being observed subsequent to $z = 0.5$, while \citet{muzzi13b} measured an increase of 1.6 times in the stellar mass density of individual $M>10^{11.5}M_{\sun}$ galaxies from $z = 2.0$ to $z = 0.3$.

\vspace{20 pt}

\section{Summary}
\label{sec:summary}

We measured evolution of the $B$-band LF from $z = 1.2$ to $z = 0.2$, improving on the prior literature by using a large galaxy sample of $408 \, 495$ selected from an  8.26 deg$^2$ field in \bootess with $uB_wRIyzJHK_s $ optical and infrared photometry from NDWFS, LBT \bootes, NEWFIRM and Subaru, together with 3.6, 4.5, 5.8 and 8.0 $\mu$m infrared photometry from SDWFS. We used a variable aperture size and corrected for flux falling outside the photometric aperture in order to accurately measure total galaxy light as a function of magnitude. Absolute magnitudes were determined using the large catalogue of 129 SED templates from \citet{brown14} and the method of \citet{beare14} which minimises the impact of systematic errors. Our photometric redshifts were based on fits to the templates of \citet{brown14}. Evolution of the Schechter parameters $M^*$ and $\phi^*$ for red, blue and all galaxies was measured for fixed $\alpha$ values of -0.5, -1.3 and -1.1 respectively, corresponding to the values in our two lowest redshift bins, i.e. $0.2 \leq z < 0.6$, when $\alpha$ was treated as a free parameter.

Our measurements were compared with those from other studies \citep{bell04, willm06, faber07, brown07, zucca09, cool12, fritz14}, and in the low redshift Universe with \citet{madgw02, blant05, loved12}. We separated ``red'' and ``blue'' galaxies using an evolving cut in restframe $(M_U - M_B)$ versus $M_B$ color-magnitude space, whereas many other authors have used a variety of other methods.

Blue galaxies are more numerous than red at all redshifts and are present in rapidly increasing numbers as one goes to fainter magnitudes (faint end slope parameter $\alpha = -1.3$). The numbers of red galaxies show a downturn and decrease rapidly at fainter magnitudes ($\alpha = -0.5$). The characteristic space density $\phi^*$ for blue $\sim L^*$ galaxies hardly changed from $z \sim 0.9$ to $z \sim 0.3$ while that of red galaxies increased by \s50\% from $z \sim 1.1$ to $z \sim 0.3$. The characteristic magnitude $M^*$ of blue galaxies faded more than that of red (0.8 as opposed to 0.6 mag per unit redshift). 

The total luminosity density $j$ of red galaxies increased marginally from $z \sim 1.1$ to $z \sim 0.3$, while that of blue galaxies almost halved from $z \sim 0.9$ to $z \sim 0.3$ (our results for blue galaxies at $1.0 \leq z < 1.2$ are uncertain because of indeterminate systematic redshift errors.) 

We included the low redshift ($z \sim 0.1$) 2dFGRS results in our analysis and compared the fading of luminosity density in our red galaxy sample with that to be expected on the basis of passive evolution alone (i.e. no star formation). From this we inferred that the stellar mass in red galaxies increased  by a factor of \s3.6 from $z = 1.1$ to $z = 0.1$.

The luminosity of individual highly luminous red galaxies decreased by 0.4 mag from $z = 1.1$ to $z = 0.3$. When low redshift ($z \sim 0.1$) results from 2dFGRS were included,  we found that the rate of fading increased from \s0.2 mag per unit redshift at $z = 1.0$ to \s0.8 at $z = 0.2$. We compared the observed fading of the bright end of the LF for red galaxies with that to be expected for a passively evolving model and concluded that individual highly luminous red galaxies increased in mass by a factor of \s2.2 from $z = 1.1$ to $z = 0.1$.

\section{ACKNOWLEDGEMENTS}
\label{sec:acknowledgements}

Richard Beare wishes to thank Monash University for financial support from MGS and MIPRS postgraduate research scholarships. Michael Brown acknowledges financial support from The Australian Research Council (FT100100280) and the Monash Research Accelerator Program (MRA). Yen-Ting Lin is grateful to Naoki Yasuda for assistance with the reduction of the Subaru $z$-band data, and to the HSC software team for developing the HSC reduction pipeline. We thank colleagues on the NDWFS, SDWFS, NEWFIRM \bootes, and AGES teams, in particular M. L. N. Ashby, R. J. Cool, A. Dey, P. R. Eisenhardt, D. J. Eisenstein, A. H. Gonzalez, B. T. Jannuzi, C. S. Kochanek and D. Stern. We are also grateful to J. Loveday and M. R. Blanton for advice on their low redshift luminosity functions, and to the anonymous referee for helpful comments which have significantly improved the paper. This work is based in part on observations made with the Spitzer Space Telescope, which is operated by the Jet Propulsion Laboratory, California Institute of Technology under a contract with NASA. This research was supported by the National Optical Astronomy Observatory, which is operated by the Association of Universities for Research in Astronomy (AURA), Inc., under a cooperative agreement with the National Science Foundation.

\begin{deluxetable*}{ccccccc}
\tablecolumns{7}
\tabletypesize{\scriptsize}
\tablecaption{$B$-band luminosity functions for all galaxies.}
\tablehead{
\multicolumn {2}{c}{$M_B - 5\log h_{70}$} & \multicolumn {5}{c}{Luminosity Function ($10^{-3} h_{70}^3 \rm{ Mpc}^{-3} \rm{ mag}^{-1}$)}\\
\colhead{Min} & \colhead{Max} &\colhead{$0.2 \leq z < 0.4 $} & \colhead{$0.4 \leq z < 0.6 $} & \colhead{$0.6 \leq z < 0.8 $} & \colhead{$0.8 \leq z < 1.0 $} & \colhead{$1.0 \leq z < 1.2 $}}
\startdata
 & & & & & & \\
$   -24.00$ & $   -23.75$ &    -     &    -                              &    -                              &  $  0.001\pm  0.001$ &      -                          \\
$   -23.75$ & $   -23.50$ &    -     & $   0.001\pm  0.001$ & $   0.001\pm  0.001$ & $   0.001\pm  0.001$ & $   0.003\pm  0.001$\\
$   -23.50$ & $   -23.25$ &    -     & $   0.001\pm  0.001$ & $   0.001\pm  0.001$ & $   0.005\pm  0.001$ & $   0.006\pm  0.001$\\
$   -23.25$ & $   -23.00$ &    -     & $   0.007\pm  0.002$ & $   0.007\pm  0.002$ & $   0.017\pm  0.002$ & $   0.024\pm  0.003$\\
$   -23.00$ & $   -22.75$ & $   0.001\pm  0.001$ & $   0.014\pm  0.003$ & $   0.027\pm  0.003$ & $   0.054\pm  0.004$ & $   0.054\pm  0.004$\\
$   -22.75$ & $   -22.50$ & $   0.004\pm  0.003$ & $   0.044\pm  0.005$ & $   0.073\pm  0.006$ & $   0.130\pm  0.007$ & $   0.122\pm  0.006$\\
$   -22.50$ & $   -22.25$ & $   0.054\pm  0.009$ & $   0.111\pm  0.009$ & $   0.172\pm  0.009$ & $   0.250\pm  0.009$ & $   0.234\pm  0.008$\\
$   -22.25$ & $   -22.00$ & $   0.143\pm  0.015$ & $   0.275\pm  0.014$ & $   0.336\pm  0.012$ & $   0.452\pm  0.012$ & $   0.408\pm  0.011$\\
$   -22.00$ & $   -21.75$ & $   0.309\pm  0.021$ & $   0.508\pm  0.019$ & $   0.563\pm  0.016$ & $   0.791\pm  0.016$ & $   0.693\pm  0.014$\\
$   -21.75$ & $   -21.50$ & $   0.496\pm  0.027$ & $   0.808\pm  0.023$ & $   0.849\pm  0.019$ & $   1.150\pm  0.020$ & $   0.986\pm  0.017$\\
$   -21.50$ & $   -21.25$ & $   0.832\pm  0.035$ & $   1.169\pm  0.028$ & $   1.234\pm  0.023$ & $   1.693\pm  0.024$ &    -    \\
$   -21.25$ & $   -21.00$ & $   1.155\pm  0.042$ & $   1.654\pm  0.034$ & $   1.627\pm  0.027$ & $   2.249\pm  0.028$ &    -    \\
$   -21.00$ & $   -20.75$ & $   1.621\pm  0.049$ & $   2.167\pm  0.038$ & $   2.152\pm  0.031$ &    -     &    -    \\
$   -20.75$ & $   -20.50$ & $   2.026\pm  0.055$ & $   2.545\pm  0.042$ & $   2.651\pm  0.035$ &    -     &    -    \\
$   -20.50$ & $   -20.25$ & $   2.428\pm  0.060$ & $   3.014\pm  0.045$ & $   2.995\pm  0.037$ &    -     &    -    \\
$   -20.25$ & $   -20.00$ & $   2.853\pm  0.065$ & $   3.479\pm  0.049$ & $   3.445\pm  0.040$ &    -     &    -    \\
$   -20.00$ & $   -19.75$ & $   3.216\pm  0.069$ & $   3.902\pm  0.052$ &    -     &    -     &    -    \\
$   -19.75$ & $   -19.50$ & $   3.976\pm  0.077$ & $   4.258\pm  0.055$ &    -     &    -     &    -    \\
$   -19.50$ & $   -19.25$ & $   4.395\pm  0.081$ &    -     &    -     &    -     &    -    \\
$   -19.25$ & $   -19.00$ & $   5.028\pm  0.087$ &    -     &    -     &    -     &    -    \\
$   -19.00$ & $   -18.75$ & $   5.513\pm  0.091$ &    -     &    -     &    -     &    -    \\
$   -18.75$ & $   -18.50$ & $   6.201\pm  0.098$ &    -     &    -     &    -     &    -    \\
\enddata
\label{tab:bin_densities_redandblue}
\end{deluxetable*}

\begin{deluxetable*}{ccccccc}
\tablecolumns{7}
\tabletypesize{\scriptsize}
\tablecaption{$B$-band luminosity functions for red galaxies.}
\tablehead{
\multicolumn {2}{c}{$M_B - 5\log h_{70}$} & \multicolumn {5}{c}{Luminosity Function ($10^{-3} h_{70}^3 \rm{ Mpc}^{-3} \rm{ mag}^{-1}$)}\\
\colhead{Min} & \colhead{Max} &\colhead{$0.2 \leq z < 0.4 $} & \colhead{$0.4 \leq z < 0.6 $} & \colhead{$0.6 \leq z < 0.8 $} & \colhead{$0.8 \leq z < 1.0 $} & \colhead{$1.0 \leq z < 1.2 $}}
\startdata
 & & & & & & \\
$   -23.50$ & $   -23.25$ &    -     &   - & - & $   0.003\pm  0.001$ & $   0.001\pm  0.001$\\
$   -23.25$ & $   -23.00$ &    -     & $   0.003\pm  0.002$ & $   0.004\pm  0.001$ & $   0.008\pm  0.002$ & $   0.010\pm  0.002$\\
$   -23.00$ & $   -22.75$ &    - & $   0.010\pm  0.003$ & $   0.016\pm  0.003$ & $   0.024\pm  0.003$ & $   0.021\pm  0.002$\\
$   -22.75$ & $   -22.50$ & $   0.001\pm  0.001$ & $   0.024\pm  0.004$ & $   0.037\pm  0.004$ & $   0.057\pm  0.004$ & $   0.047\pm  0.004$\\
$   -22.50$ & $   -22.25$ & $   0.037\pm  0.007$ & $   0.069\pm  0.007$ & $   0.082\pm  0.006$ & $   0.118\pm  0.006$ & $   0.091\pm  0.005$\\
$   -22.25$ & $   -22.00$ & $   0.093\pm  0.012$ & $   0.152\pm  0.010$ & $   0.147\pm  0.008$ & $   0.198\pm  0.008$ & $   0.167\pm  0.007$\\
$   -22.00$ & $   -21.75$ & $   0.173\pm  0.016$ & $   0.270\pm  0.014$ & $   0.237\pm  0.010$ & $   0.321\pm  0.010$ & $   0.269\pm  0.009$\\
$   -21.75$ & $   -21.50$ & $   0.250\pm  0.019$ & $   0.398\pm  0.016$ & $   0.350\pm  0.012$ & $   0.465\pm  0.013$ & $   0.377\pm  0.011$\\
$   -21.50$ & $   -21.25$ & $   0.430\pm  0.025$ & $   0.532\pm  0.019$ & $   0.473\pm  0.014$ & $   0.605\pm  0.014$ &    -    \\
$   -21.25$ & $   -21.00$ & $   0.613\pm  0.030$ & $   0.759\pm  0.023$ & $   0.562\pm  0.016$ & $   0.720\pm  0.016$ &    -    \\
$   -21.00$ & $   -20.75$ & $   0.752\pm  0.033$ & $   0.901\pm  0.025$ & $   0.654\pm  0.017$ & $   0.871\pm  0.018$ &    -    \\
$   -20.75$ & $   -20.50$ & $   0.813\pm  0.035$ & $   0.986\pm  0.026$ & $   0.747\pm  0.018$ &    -     &    -    \\
$   -20.50$ & $   -20.25$ & $   0.950\pm  0.038$ & $   1.022\pm  0.026$ & $   0.701\pm  0.018$ &    -     &    -    \\
$   -20.25$ & $   -20.00$ & $   1.022\pm  0.039$ & $   1.123\pm  0.028$ & $   0.754\pm  0.019$ &    -     &    -    \\
$   -20.00$ & $   -19.75$ & $   1.018\pm  0.039$ & $   1.151\pm  0.028$ & $   0.742\pm  0.019$ &    -     &    -    \\
$   -19.75$ & $   -19.50$ & $   1.018\pm  0.039$ & $   1.102\pm  0.028$ &    -     &    -     &    -    \\
$   -19.50$ & $   -19.25$ & $   1.040\pm  0.039$ & $   1.138\pm  0.028$ &    -     &    -     &    -    \\
$   -19.25$ & $   -19.00$ & $   1.051\pm  0.040$ & $   1.002\pm  0.027$ &    -     &    -     &    -    \\
$   -19.00$ & $   -18.75$ & $   0.934\pm  0.037$ & $   0.876\pm  0.025$ &    -     &    -     &    -    \\
$   -18.75$ & $   -18.50$ & $   0.927\pm  0.037$ &    -     &    -     &    -     &    -    \\
$   -18.50$ & $   -18.25$ & $   0.882\pm  0.037$ &    -     &    -     &    -     &    -    \\
$   -18.25$ & $   -18.00$ & $   0.910\pm  0.037$ &    -     &    -     &    -     &    -    \\
$   -18.00$ & $   -17.75$ & $   0.895\pm  0.037$ &    -     &    -     &    -     &    -    \\
$   -17.75$ & $   -17.50$ & $   0.862\pm  0.037$ &    -     &    -     &    -     &    -    \\
\enddata
\label{tab:bin_densities_red}
\end{deluxetable*}

\begin{deluxetable*}{ccccccc}
\tablecolumns{7}
\tabletypesize{\scriptsize}
\tablecaption{$B$-band luminosity functions for blue galaxies.}
\tablehead{
\multicolumn {2}{c}{$M_B - 5\log h_{70}$} & \multicolumn {5}{c}{Luminosity Function ($10^{-3} h_{70}^3 \rm{ Mpc}^{-3} \rm{ mag}^{-1}$)}\\
\colhead{Min} & \colhead{Max} &\colhead{$0.2 \leq z < 0.4 $} & \colhead{$0.4 \leq z < 0.6 $} & \colhead{$0.6 \leq z < 0.8 $} & \colhead{$0.8 \leq z < 1.0 $} & \colhead{$1.0 \leq z < 1.2 $}}
\startdata
 & & & & & & \\
$   -23.75$ & $   -23.50$ &    -     & $   0.001\pm  0.001$ & $   0.001\pm  0.001$ & $   0.001\pm  0.001$ & $   0.002\pm  0.001$\\
$   -23.50$ & $   -23.25$ &    -     & $   0.001\pm  0.001$ & $   0.001\pm  0.001$ & $   0.002\pm  0.001$ & $   0.004\pm  0.001$\\
$   -23.25$ & $   -23.00$ &    -     & $   0.004\pm  0.002$ & $   0.003\pm  0.001$ & $   0.009\pm  0.002$ & $   0.015\pm  0.002$\\
$   -23.00$ & $   -22.75$ &    -     & $   0.004\pm  0.002$ & $   0.011\pm  0.002$ & $   0.030\pm  0.003$ & $   0.034\pm  0.003$\\
$   -22.75$ & $   -22.50$ & $   0.003\pm  0.002$ & $   0.020\pm  0.004$ & $   0.037\pm  0.004$ & $   0.072\pm  0.005$ & $   0.075\pm  0.005$\\
$   -22.50$ & $   -22.25$ & $   0.016\pm  0.005$ & $   0.042\pm  0.005$ & $   0.090\pm  0.006$ & $   0.132\pm  0.007$ & $   0.143\pm  0.006$\\
$   -22.25$ & $   -22.00$ & $   0.051\pm  0.009$ & $   0.123\pm  0.009$ & $   0.189\pm  0.009$ & $   0.254\pm  0.009$ & $   0.240\pm  0.008$\\
$   -22.00$ & $   -21.75$ & $   0.136\pm  0.014$ & $   0.237\pm  0.013$ & $   0.326\pm  0.012$ & $   0.470\pm  0.013$ & $   0.423\pm  0.011$\\
$   -21.75$ & $   -21.50$ & $   0.247\pm  0.019$ & $   0.410\pm  0.017$ & $   0.499\pm  0.015$ & $   0.685\pm  0.015$ & $   0.608\pm  0.013$\\
$   -21.50$ & $   -21.25$ & $   0.402\pm  0.024$ & $   0.637\pm  0.021$ & $   0.761\pm  0.018$ & $   1.088\pm  0.019$ & $   0.954\pm  0.017$\\
$   -21.25$ & $   -21.00$ & $   0.542\pm  0.028$ & $   0.895\pm  0.025$ & $   1.065\pm  0.022$ & $   1.529\pm  0.023$ &    -    \\
$   -21.00$ & $   -20.75$ & $   0.870\pm  0.036$ & $   1.266\pm  0.029$ & $   1.498\pm  0.026$ &    -     &    -    \\
$   -20.75$ & $   -20.50$ & $   1.213\pm  0.043$ & $   1.559\pm  0.033$ & $   1.904\pm  0.029$ &    -     &    -    \\
$   -20.50$ & $   -20.25$ & $   1.478\pm  0.047$ & $   1.992\pm  0.037$ & $   2.295\pm  0.033$ &    -     &    -    \\
$   -20.25$ & $   -20.00$ & $   1.831\pm  0.052$ & $   2.355\pm  0.040$ & $   2.691\pm  0.035$ &    -     &    -    \\
$   -20.00$ & $   -19.75$ & $   2.198\pm  0.057$ & $   2.751\pm  0.044$ &    -     &    -     &    -    \\
$   -19.75$ & $   -19.50$ & $   2.958\pm  0.066$ & $   3.156\pm  0.047$ &    -     &    -     &    -    \\
$   -19.50$ & $   -19.25$ & $   3.355\pm  0.071$ &    -     &    -     &    -     &    -    \\
$   -19.25$ & $   -19.00$ & $   3.977\pm  0.077$ &    -     &    -     &    -     &    -    \\
$   -19.00$ & $   -18.75$ & $   4.579\pm  0.083$ &    -     &    -     &    -     &    -    \\
\enddata
\label{tab:bin_densities_blue}
\end{deluxetable*}


\bibliographystyle{apj}

\end{document}